\documentclass{elsarticle} \usepackage{url} \usepackage{xcolor}
\definecolor{newcolor}{rgb}{.8,.349,.1}

\usepackage[utf8]{inputenc} \usepackage{graphicx} \usepackage{comment}
\usepackage{natbib}
\usepackage{amsmath} \usepackage{latexsym} \usepackage{amssymb}
\usepackage{mathrsfs} \usepackage{comment, url}
\usepackage{caption,subcaption} \usepackage{enumitem}

\definecolor{MyBlue}{rgb} {0.1,0.1,0.9} \definecolor{MyRed}{rgb}
{0.9,0.1,0.1}
\definecolor{MyGreen}{rgb} {0.05,0.4,0.05}

\usepackage[bbgreekl]{mathbbol} \usepackage{units}

\newcommand{\rr}{\mathbf{r}} \newcommand{\Id}{\mathbf{I}}
\newcommand{\xx}{\mathbf{x}} \newcommand{\vv}{\mathbf{v}}
\newcommand{\pp}{\mathbf{p}} \newcommand{\eps}{\varepsilon}
\newcommand{\LL}{\mathbf{L}} \newcommand{\XX}{\mathbf{X}}
\newcommand{\YY}{\mathbf{Y}} \newcommand{\mm}{\mathbf{m}}
\newcommand{\R}{\mathbb{R}} \newcommand{\GGG}{\mathfrak{g}}
\newcommand{\oomega}{\boldsymbol{\omega}}

\newcommand{\GGamma}{\boldsymbol{\Gamma}} \newcommand{\uu}{\mathbf{u}}
\newcommand{\dr}{\mathrm{d}\rr} \newcommand{\dpp}{\mathrm{d}\pp}
\newcommand{\diff}{\mathrm{d}} \newcommand{\Lie}{\mathcal{L}}
\newcommand{\Obig}{\mathcal{O}} \newcommand{\Par}{\mathcal{P}}
\newcommand{\mmu}{{\boldsymbol{\mu}}} \newcommand{\rev}{\mathrm{rev}}
\newcommand{\MM}{\mathbf{M}} \newcommand{\NN}{\mathbf{N}}
\newcommand{\AAA}{\mathcal{A}} \newcommand{\UUU}{\mathcal{U}}

\begin{document}

\title{Ehrenfest regularization of Hamiltonian systems}
\author[mff]{Michal Pavelka\corref{cor1}}
\ead{pavelka@karlin.mff.cuni.cz}
\author[fjfi]{Václav Klika}
\cortext[cor1]{Corresponding author}
\address[mff]{Mathematical Institute, Faculty of Mathematics and Physics, Charles University, Sokolovská 83, 186 75 Prague, Czech Republic}
\address[fjfi]{Department of Mathematics, FNSPE, Czech Technical University in Prague, Trojanova 13, Prague 2, 120 00, Czech Republic}
\author[poly]{Miroslav Grmela}
\address[poly]{\'{E}cole Polytechnique de Montr\'{e}al, C.P.6079 suc. Centre-ville, Montr\'{e}al, H3C 3A7,  Qu\'{e}bec, Canada}

\begin{abstract}
	Imagine a freely rotating rigid body. The body has three principal axes of rotation. It follows from mathematical analysis of the evolution equations that pure rotations around the major and minor axes are stable while rotation around the middle axis is unstable. However, only rotation around the major axis (with highest moment of inertia) is stable in physical reality (as demonstrated by the unexpected change of rotation of the Explorer 1 probe). We propose a general method of Ehrenfest regularization of Hamiltonian equations by which the reversible Hamiltonian equations are equipped with irreversible terms constructed from the Hamiltonian dynamics itself. The method is demonstrated on harmonic oscillator, rigid body motion (solving the problem of stable minor axis rotation), ideal fluid mechanics and kinetic theory. In particular, the regularization can be seen as a birth of irreversibility and dissipation.  In addition, we discuss and propose discretizations of the Ehrenfest regularized evolution equations such that key model characteristics (behavior of energy and entropy) are valid in the numerical scheme as well.
\end{abstract}


\maketitle

\numberwithin{equation}{section}

\section{Introduction}
Consider an isolated  system  whose time evolution is governed by Hamiltonian mechanics. One can solve the Hamilton evolution equations and see how the system evolves in time. In reality, however, the system seldom evolves just according to the reversible dynamics from the long time perspective, for not being completely isolated (either due to interaction with its surroundings or even with another degrees of freedom of the physical system itself), for exhibiting also another kinds of motion or evolution, or by the loss of regularity of the solutions and damping of fast oscillations.

In this paper we put into focus  the  irregularities in solutions. We are not investigating  their emergence (as it is done for example in \cite{landau,Villani2014,PKG-Landau}) but  their smoothing and transforming into regular solutions of a modified (regularized) time evolution equations.
There is no, we believe, universal way of "smoothing out". In this manuscript we smooth out by passing from the vector field to the trajectory (but only a very short trajectory).  In other words, the physics of the proposed regularization is the passage from the vector field to the generated by it trajectory (but only an infinitesimal step further). From the physical point of view again, the essence in getting something "overall" out of solutions is to recognize pattern in the phase portrait (collection of all trajectories i.e. solutions). To get the whole trajectory is, of course, a very difficult problem. So let us be satisfied with at least a very short trajectory while being compatible with thermodynamics. 

Another way of smoothing out is for instance explored in \cite{KRM}. The system is first enlarged by taking into account more microscopic details. The smoothing mechanism, introduced in the time evolution of the details, then trickles down to the whole system. Alternatively, the same  strategy can also be used in open systems where the enlargement would consist of inclusion of the environment and the smoothing mechanism would appear in the time evolution generated by the system-environment interactions.

The Hamiltonian evolution equations are enriched in this paper by adding Ehrenfest regularization (EhRe) terms that drive the system towards the physically meaningful states. The EhRe terms, however, are constructed in such a way that no more knowledge than that of the Hamiltonian description itself (except a relaxation time parameter) is needed.
The Ehrenfest regularization (EhRe) is shown to reduce kinetic energy (convex generator of the Hamiltonian dynamics) and produce kinetic entropy (concave Casimir of the Poisson bracket). It can be moreover split into two scenarios: (i) the energetic Ehrenfest regularization (E-EhRe) and (ii) entropic Ehrenfest regularization (S-EhRe). The E-EhRe conserves kinetic entropy while reducing kinetic energy. The S-EhRe, on the other hand, conserves kinetic energy while producing kinetic entropy. One can choose the regularization based on the particular conservation laws the system at hand obeys.

For instance, the E-EhRe is useful in rigid body dynamics, where energy of rotation is dissipated to internal energy while conserving angular momentum (a Casimir of the bracket, i.e. kinetic entropy). By adding an internal entropy (advected by the Hamiltonian dynamics), conservation of total energy can be restored while keeping conservation of the Casimirs (e.g. angular momentum) and the second law of thermodynamics (positive entropy production) appears as a consequence. The E-EhRe of the solid body rotation indicates the loss of stability of the rotation around the (minor) axis with lowest moment of inertia, and rotation around the (major) axis with highest moment of inertia becomes the sole attractor of the dynamics. That is the anticipated physical behavior \cite{Krechetnikov-Marsden}.

Next, we apply the EhRe regularization method to more complex situations. Namely, Euler equations for ideal fluids, described by state variables density and momentum density, can also be Ehrenfest-regularized. The EhRe evolution with an internal entropy (advected by the semidirect product) then leads to a system of equations transforming kinetic energy to internal energy while raising internal entropy. Further, Hamilton canonical equations for a harmonic oscillator are also regularized by the E-EhRe method, since kinetic energy is then dissipated to internal energy. Such dissipation is then accompanied by growth of internal entropy. Finally, the S-EhRe method is useful in, for instance, kinetic theory, since it makes the Boltzmann entropy grow while keeping energy of the system.

Apart from the evolution equations, it is interesting to discuss the possible numerical schemes solving them. We reveal that the full EhRe evolution can be solved by explicit Euler method with relaxation time larger than the discretized time increment, i.e. $\tau \geq \diff t$. In particular, if $\diff t = \tau$, the scheme gives an accurate explicit method for integration of the original Hamiltonian equations conserving both kinetic energy and kinetic entropy to the second order in $\diff t$ as the chosen numerical scheme. For $\tau>\diff t$ the numerical scheme becomes dissipative.

The E-EhRe and S-EhRe equations are best discretized using Crank-Nicolson (semi-implicit), since then the properties of the regularizations (conservation of entropy and reduction of energy or conservation of energy and growth of entropy) are retained by the numerical schemes.

In summary, the Ehrenfest regularization (either full, energetic or entropic) is a novel useful tool explicitly manifesting some overall damping and dissipative features of the original Hamiltonian dynamics.
The proposed regularization (Ehrenfest) approach is such that the leading order damping being explicitly included in the evolution is compatible (to the second order in $\tau$) with the (local) trajectories although they themselves might be stemming from reversible dynamics. From the physical point of view, the idea is to pass from a vector field to the corresponding to it phase portrait (i.e. collection of all trajectories generated by the vector field), recognize in it an overall pattern, and subsequently to identify a new, modified, vector field generating the pattern. Admittedly, we have not provided a proof that the explicit damping terms correspond exactly to the emergent properties of reversible dynamics (as in Landau damping shown by \cite{Villani2014}) but the universality of the method together with its construction (physical motivation) makes us believe that it is worth pursuing. 
It can be used to construct accurate numerical schemes that either (i) reduce (convex) kinetic energy and keep kinetic entropy (Casimirs) or (ii) raise kinetic entropy (concave Casimirs) and keep energy (iii) reduce energy and produce entropy or (iv) keep both energy and entropy constant. The regularizations are demonstrated on several examples including their numerical solutions.

\section{Hamiltonian dynamics}
Let us first briefly recall what Hamiltonian dynamics means. Further details and connections to thermodynamics can be found for instance in book \cite{PKG}.

\subsection{Poisson bracket and bivector}
Consider system described by state variables $\xx$ evolution of which is Hamiltonian. The evolution equations for the state variables are then
\begin{equation}\label{eq.Ham}
	\dot{x}^i = L^{ij} E_{x^j},
\end{equation}
where $L^{ij} = -L^{ji}$ is an Poisson bivector (antisymmetric twice contravariant tensor satisfying Jacobi identity \cite{Fecko}) and $E_{x^j}$ are derivatives of an energy functional (Hamiltonian) also called kinetic energy in this paper. The right hand side can also be seen as components of a Hamiltonian vector field. The Poisson bivector generates Poisson bracket
\begin{equation}
	\{F,G\} =  F_{x^i} L^{ij}G_{x^j},
\end{equation}
which is a bilinear antisymmetric operator, $\{F,G\}=-\{G,F\}$, fulfilling Jacobi identity,
\begin{equation}
	0 = \{F, \{G, H\}\}+ \{G, \{H, F\}\}+ \{H, \{F, G\}\}.
\end{equation}
Substituting energy $E$ for $G$ leads to a weak formulation of the dynamics
\begin{equation}
	\{F,E\} = F_{x^i} \dot{x}^i = \dot{F}.
\end{equation}
Evolution of an arbitrary functional $F$ is thus generated by the Poisson bracket and energy.

The Hamiltonian evolution equations are reversible in two senses. First, they are invariant with respect to the time-reversal transformation (TRT), see e.g. \cite{pre2014,PKG}. Secondly, since entropy (here called kinetic entropy) is always assumed to be a Casimir of the Poisson bracket,
\begin{equation}
\{S,G\} = 0 \quad\forall G, \qquad \iff L^{ij}S_{x^j} = 0 \quad\forall\, i,
\end{equation}
entropy is not affected by the Hamiltonian dynamics. Hamiltonian dynamics is reversible with respect to the TRT and does not change (kinetic) entropy.

\subsection{Stability by the Energy-Casimir method}\label{sec.ECM}
The purpose of this section is to recall the Energy-Casimir method \cite{Holm-stability}, which is a useful tool for proving nonlinear stability of stationary states of Hamiltonian systems. The method is a variant of Arnold's Lyapunov method \cite{Arnold1969}, see \cite{Holm-stability} for historical overview.

Consider a set of state variables and their evolution equations (not necessarily Hamiltonian)
\begin{equation}
	\dot{x}^i = X^i(\xx).
\end{equation}
The right hand side can be for instance a Hamiltonian vector field, Eq. \eqref{eq.Ham}. Let a point $\xx_0$ be a stationary point of the evolution, i.e. $\dot{x}^i_0 = 0 = X^i(\xx_0)$. Assume, moreover, that energy $E(\xx)$ is constant along the evolution and that there is a functional $S(\xx)$ that is constant along the evolution as well. In the case of Hamiltonian dynamics energy is conserved automatically due to the antisymmetry of the Poisson bracket and entropy $S$ is a Casimir of the underlying Poisson bracket and is thus also conserved. Note that there is usually a whole class of Casimirs of the Poisson bracket, so there is a whole class of conserved quantities that can be used as (kinetic) entropy $S$.

The algorithm of the Energy-Casimir method (see e.g. \cite{Holm-stability} for an accessible and comprehensive discussion) consists of the following steps:
\begin{itemize}
\item Choose an integral of motion $S$ and a constant $\alpha$ so that the stationary point $\xx_0$ becomes also a critical point of potential $\Phi$
  \begin{equation}\label{eq.ECM.Phi}
	  \Phi = -\alpha S + E, \qquad \Phi_\xx|_{\xx_0} = 0.
		\end{equation}
Hence the potential has zero gradient at $\xx_0$, and it is a conserved quantity,
		\begin{equation}
			\dot{\Phi} = 0,
		\end{equation}
due to its construction.
\item Evaluate the second variation of $\Phi$ at $\xx_0$,
	\begin{equation}
			\delta^2 \Phi|_{\xx_0} = \delta x^i \frac{\partial^2 \Phi}{\partial x^i \partial x^j}\Big|_{\xx_0}\delta x^j.
		\end{equation}
There are now several possibilities:\footnote{
  Note that for infinite-dimensional systems the criterion of definiteness of the second variation should be replaced with convexity or concavity of $\Phi$.}
  \begin{itemize}
	  \item Strict convexity near $\xx_0$, i.e. $\delta^2 \Phi > 0\quad \forall \delta\xx\in\UUU(\xx_0)$ from a neighborhood of $\xx_0$, which means that there is a strictly convex quadratic form $Q(\delta\xx)$ such that\footnote{Note that similar potential is used for showing Lyapunov stability of stationary non-equilibrium states \cite{Vitek-Lyapunov}.}
  \begin{equation}
					0 \leq Q(\delta \xx) \leq \Phi(\xx_0+\delta\xx) - \left(\Phi(\xx_0) + \Phi_{x^i}|_{\xx_0} \delta x^i\right) = \Phi(\xx_0+\delta\xx)-\Phi(\xx_0)
				\end{equation}
  while the equality $Q(\delta\xx)=0$ be fulfilled only for $\delta\xx =0 $.
  \item Strict concavity near $\xx_0$, i.e. $\delta^2 \Phi < 0\quad \forall \delta\xx$, which means that there is a strictly quadratic form $Q(\delta\xx)$ such that
  \begin{equation}
					0 \leq Q(\delta \xx) \leq -\left[\Phi(\xx_0+\delta\xx) - \left(\Phi(\xx_0) + \Phi_{x^i}|_{\xx_0} \delta x^i\right)\right] = -\left[\Phi(\xx_0+\delta\xx)-\Phi(\xx_0)\right],
				\end{equation}
  while $Q(\delta \xx)=0$ only for $\delta\xx = 0$.
  \item Indefiniteness of $\delta^2\Phi$, which is an indication (though not a proof) of instability of state $\xx_0$.
 \end{itemize}
\item For the first two options (positive or negative definite case), the quadratic form $Q$ defines a norm on the space of state variables
		\begin{equation}
			0\leq ||\xx-\xx_0||_Q \stackrel{\mathrm{def}}{=} Q(\xx-\xx_0) \leq \pm(\Phi(\xx)-\Phi(\xx_0)),
		\end{equation}
which is bounded from above due to conservative nature of the potential $\Phi$.
In addition, as both integrals of motion are typically continuously dependent on initial data and since the right hand side is a conserved quantity, we can conclude that for any $\epsilon>0$ there is a neighborhood of $\xx_0$ for which $|\Phi(\xx)-\Phi(\xx_0)|\leq \epsilon$ and such that if the evolution starts in the neighborhood, it remains there and thus $||\xx-\xx_0||_Q\leq \epsilon$ $\forall t>0$. The point $\xx_0$ is then stable in this sense.
\end{itemize}
The Energy-Casimir method can be used to prove nonlinear stability of stationary points of the evolution, and which is especially useful for Hamiltonian systems due to the inherent presence of Casimirs. It will be demonstrated on the rigid body motion in Sec. \ref{sec.SO3.stab}.

\subsection{Formal solution of Hamiltonian dynamics}
The right hand side of evolution equations \eqref{eq.Ham} can also be regarded as the Hamiltonian vector field with components
\begin{equation}
	X^i_E  = L^{ij} E_{x^j},
\end{equation}
which can be expressed as
\begin{equation}
	\frac{\diff \xx(t)}{\diff t}  = \Lie_{X_E} \xx(t),
\end{equation}
$\Lie_{X_E}$ being the Lie derivative with respect to the Hamiltonian vector field.
Note that the time-derivative is interpreted as partial time-derivative in the case of partial differential equations and the summation is interpreted as integration over the spatial domain in that case.
Formal solution to this equations reads (see e.g. \cite{Fecko} or \cite{PKG})
\begin{eqnarray}
	\xx(t+\tau) &=& \exp\left(\tau \Lie_{X_E}\right)\xx(t) = \xx(t) + \tau \Lie_{X_E} \xx(t) + \frac{\tau^2}{2} \Lie_{X_E}\Lie_{X_E} \xx(t) + \Obig(\tau)^3.
\end{eqnarray}
In other words, at time $t+\tau$ we have
\begin{equation}\label{eq.xtau}
	x^i(t+\tau) = x^i(t) + \tau L^{ij} E_{x^j} + \frac{\tau^2}{2} \frac{\delta}{\delta x^k}\left(L^{ij}E_{x^j}\right)L^{kl}E_{x^l} + \Obig(\tau^3),
\end{equation}
which is a formal solution of Eq. \eqref{eq.Ham} for time-step $\tau$ hence describing local trajectories (phase portrait). This last equation can also be rewritten as
\begin{equation}\label{eq.xtau.PB}
	x^i(t+\tau) = x^i(t) + \tau L^{ij} E_{x^j} + \frac{\tau^2}{2} \{ L^{ij}E_{x^j}, E\} + \Obig(\tau^3),
\end{equation}
where $\{\bullet,\bullet\}$ is the Poisson bracket corresponding to the Poisson bivector $\LL$.

\subsection{Numerical scheme}

Equation \eqref{eq.xtau} can be used as a numerical scheme of second-order solving evolution equation \eqref{eq.Ham}. We shall now discuss properties of this scheme. Since the scheme is of second order by construction, quantities conserved by the original Hamiltonian evolution are conserved to the second order by the scheme as well. Energy at time $t+\tau$ is equal to
\begin{eqnarray} \label{Eq.18}
	E(\xx(t+\tau))&=& E(\xx(t)) + \Obig(\tau)^3,
\end{eqnarray}
which means that it is conserved up to the order $\Obig(\tau^2)$.

Entropy is required to be a Casimir of the Poisson bracket,
\begin{equation}
	\{S, G\} = 0 \quad\forall G\qquad\mbox{or}\qquad S_{x^i}L^{ij} = 0\quad\forall i.
\end{equation}
Using scheme \eqref{eq.xtau},
entropy is given at time $t+\tau$ by
\begin{eqnarray} \label{Eq.19}
	S(\xx(t+\tau))&=&  S(\xx(t)) +\Obig(\tau)^3
\end{eqnarray}
as in the case of energy. 

Prescription \eqref{eq.xtau} provides a numerical scheme (taking time step $\diff t=\tau$) that conserves both energy and entropy up to the order $\Obig(\tau^2)$. The scheme is demonstrated on rigid body rotations in Sec. \ref{sec.SO3.EhRe.num}.

\section{Hamiltonian Ehrenfest-regularized dynamics}\label{sec.SR}

Hamiltonian dynamics is fully reversible both with respect to the time-reversal transformation (TRT) and with respect to entropy, which is kept constant, \cite{pre2014}. Let us now introduce a (we believe simple and advantageous) way of adding irreversible terms to the Hamiltonian dynamics.

\subsection{Ehrenfest regularization of the evolution equations}
Equation \eqref{eq.xtau} can be interpreted as the smoothed out time change of $\xx$,
\begin{equation}\label{eq.x.evo.sr}
	\langle \dot{x}^i\rangle \stackrel{\mbox{def}}{=}\frac{1}{\tau}\int_0^\tau \dot{x}^i \diff t
	=
	\frac{x^i(t+\tau)-x^i(t)}{\tau} 
	\approx L^{ij} E_{x^j}
	+\frac{\tau}{2} \frac{\delta}{\delta x^k}\left(L^{ij}E_{x^j}\right)L^{kl}E_{x^l}.
\end{equation}
As we can see variable $\xx$ effectively undergoes Hamiltonian reversible evolution (the first term on the right hand side) and irreversible evolution (the second term) when smoothed out over time interval of length $\tau$, a constant parameter $\tau$ called \textbf{relaxation time} \footnote{Geometric version of EhRe is presented in \ref{sec.lift} employing complete tangent lift of the Hamiltonian dynamics.}. This is the \textbf{Hamiltonian Ehrenfest regularization} (EhRe). In other words, if we assign to the time derivative its average (over $\tau$), the dynamics keeps the original Hamiltonian term while a new ``overall'' dynamics emerges.

Hence the proposed regularization (Ehrenfest) approach is such that the leading order damping being explicitly included in the evolution is compatible (to the second order in $\tau$) with the (local) trajectories although they themselves might be stemming from reversible dynamics. From the physical point of view, the idea is to pass from a vector field to the corresponding to it phase portrait (i.e. collection of all trajectories generated by the vector field), recognize in it an overall pattern, and subsequently to identify a new, modified, vector field generating the pattern. 

\subsection{Irreversibility}
Let us now show that the second term in Eq. \eqref{eq.x.evo.sr} indeed produces irreversible evolution in the sense of time-reversal transformation (TRT), see e.g. \cite{pre2014}. Assuming that all state variables have definite parities $\Par(x^i)=\pm 1$, parity of the Poisson bivector $\LL$ is necessarily
\begin{equation}
	\Par(L^{ij}) = -\Par(x^i)\Par(x^j),
\end{equation}
while parity of energy is equal to one, i.e. energy is even with respect to time-reversal. TRT applied to Eq. \eqref{eq.x.evo.sr} gives
\begin{multline}
  \Par(x^i) \frac{\diff x^i}{\diff (-t')} = -\Par(x^i)\Par(x^j)L^{ij} \Par(x^j)E_{x^j} +\\
	\frac{\tau}{2} \Par(x^k)\frac{\delta}{\delta x^k}\left(-\Par(x^i)\Par(x^j)L^{ij} \Par(x^j)E_{x^j}\right)
	\cdot\left(-\Par(x^k)\Par(x^l)L^{kl}\right)\Par(x^l)E_{x^l},
\end{multline}
$t'$ being time going backwards. This last equation can be rewritten as
\begin{equation}
	\frac{\diff x^i}{\diff t'}
	=L^{ij} E_{x^j}
	-\frac{\tau}{2} \frac{\delta}{\delta x^k}\left(L^{ij}E_{x^j}\right)L^{kl}E_{x^l}.
\end{equation}
The first term on the right hand side of this equation is the same as in Eq. \eqref{eq.x.evo.sr} while the second term has the opposite sign. The first (Hamiltonian) term therefore generates reversible evolution while the second irreversible \cite{pre2014}. Note that the parameter $\tau$ is not affected by TRT, since it is assumed to be a constant (like viscosity). Hamiltonian Ehrenfest regularization thus introduces an irreversible term into the originally reversible evolution equations.

\subsection{Dissipativity}
It has been shown that the regularizing term produces irreversible evolution, i.e. evolution changing its direction under the time reversal transformation. Usually, but not always (see e.g. \cite{hco}), irreversible evolution is also dissipative in the sense that it causes entropy growth or dissipation of energy. Let us now verify that it is indeed the case of Ehrenfest regularization.

Energy is clearly conserved by the Hamiltonian part of evolution equation \eqref{eq.x.evo.sr} (the first term on the right hand side). The irreversible term alters energy as follows
\begin{equation}
	\dot{E} = \frac{\tau}{2} E_{x^i}\frac{\delta L^{ij}E_{x^j}}{\delta x^k} L^{kl} E_{x^l}
	= -\frac{\tau}{2} \underbrace{L^{ji} E_{x^i} E_{x^j x^k} L^{kl} E_{x^l}}_{\geq 0}
	+ \frac{\tau}{2} \underbrace{E_{x^i}\frac{\delta L^{ij}}{\delta x^k} E_{x^j} L^{kl} E_{x^l}}_{=0},
\end{equation}
where the first term reduces energy due to the assumed convexity of energy (and thus positive semi definiteness of the second differential of energy). The second term is zero 
due to the simultaneous symmetry and antisymmetry with respect to swapping $i$ and $j$. Therefore, we obtain that
\begin{equation}\label{eq.SR.dE}
	\dot{E} =  -\frac{\tau}{2} L^{ji} E_{x^i} E_{x^j x^k} L^{kl} E_{x^l}\leq 0.
\end{equation}
Energy is dissipated by the EhRe evolution.

Entropy is clearly conserved by the Hamiltonian (reversible) part of the regularized evolution, since it is assumed to be a Casimir of the underlying Poisson bracket. The irreversible part of the EhRe evolution changes entropy as follows
\begin{eqnarray}\label{eq.SR.dS}
	\dot{S} &=& \frac{\tau}{2} S_{x^i} \frac{\delta L^{ij} E_{x^j}}{\delta x^k}L^{kl}E_{x^l} \nonumber\\
	&=&\frac{\tau}{2} \frac{\delta}{\delta x^k}\left(\underbrace{S_{x^i}L^{ij}}_{=0}E_{x^j}\right) L^{kl}E_{x^l}
	-\frac{\tau}{2} S_{x^i x^k} L^{ij}E_{x^j} L^{kl} E_{x^l}\geq 0.
\end{eqnarray}
The inequality follows from the additional requirement that entropy be concave, which means that the second differential of entropy is negative semidefinite. The EhRe evolution thus produces entropy provided entropy is a concave Casimir of the underlying Poisson bracket. In short, it reduces (kinetic) energy and produces (kinetic) entropy.

\subsection{Energy-Casimir method with dissipation}\label{sec.ECM.dis}
The Energy-Casimir method, which was recalled in Sec. \ref{sec.ECM}, is a convenient tool for showing stability of stationary states of Hamiltonian evolution, which is fully reversible. The regularized Hamiltonian evolution as proposed in Section ~\ref{sec.SR}, on the other hand, contains irreversible terms, and is dissipative in the sense that (kinetic) entropy grows, $\dot{S} \geq 0$ and energy decreases $\dot{E}\leq 0$. Can the Energy-Casimir method be used also for the regularized evolution?

Assume again that a point $\xx_0$ is a stationary point of the evolution and that $\Phi_\xx|_{\xx_0}=0$,
where $\Phi$ is the potential constructed from entropy and energy as in Eq. \eqref{eq.ECM.Phi}. Taking Eq. \eqref{eq.SR.dE} and Eq. \eqref{eq.SR.dS}, the potential $\Phi$ changes in time as
\begin{equation}
	\dot{\Phi} = -\alpha \dot{S} + \dot{E}
	= -\frac{\tau}{2}\left(L^{ij}E_{x^j}\right) \Phi_{x^ix^k} \left(L^{kl}E_{x^l}\right),
\end{equation}
which is negative, $\dot\Phi\leq0$, for $\Phi$ convex while positive, $\dot\Phi\geq0$ for $\Phi$ concave.

If $\Phi$ is strictly convex in a neighborhood of $\xx_0$, we can again construct the quadratic form $0\leq Q\leq \Phi(\xx)-\Phi(\xx_0)$ as in Sec. \ref{sec.ECM}. This time, however, the potential decreases in time, $\dot{\Phi}\leq 0$, as EhRe increases concave Casimirs and only those were considered in the stability analysis of Hamiltonian systems. It means that the stationary point is stable, and even asymptotically stable if it is the only point in the neighborhood where $\dot{\Phi}$ vanishes (where $L^{ij}E_{x^j}$ vanishes). Denoting the maximal connected set of all points where $\dot{\Phi}=0$ containing $\xx_0$ as the attractor $\AAA(\xx_0)$, the evolution stays in the neighborhood $\UUU(\xx_0)$ of $\xx_0$ while converging to some point in $\AAA(\xx_0)\cap \UUU(\xx_0)$. Let us refer to the notion of stability as a \textbf{nearly asymptotic stability}.

If, on the other hand, the potential is strictly concave near $\xx_0$, we have the inequality $Q(\delta\xx)\leq -\Phi(\xx) + \Phi(\xx_0)$. The potential $\Phi$ grows until it approaches the point $\xx_0$, where the right hand side becomes zero. The stationary point $\xx_0$ is thus again stable, and it is even asymptotically stable if it is the only point on the neighborhood where $\dot{\Phi}$ vanishes. The point is moreover nearly asymptotically stable in the sense above.

In summary, if stable points are found for the reversible (Hamiltonian) evolution, they are stable also with respect to the EhRe evolution. Moreover, EhRe evolution makes the stable points also nearly asymptotically stable. Note that this is exactly in line with the motivation we had for EhRe regularization - the EhRe was such a regularization that the qualitative properties of the original evolution are attained in the regularized version while being more evident, explicit in the latter.  Finally, we do not have equivalence between stability in the original Hamiltonian evolution and stability in the regularized dynamics (which would be ideal). However, we at least know one implication which is another reason why we think the proposed regularization method is worthy studying.

\subsection{Numerical scheme}
Ehrenfest regularized evolution equation \eqref{eq.x.evo.sr} can be discretized by simple forward Euler scheme,
\begin{equation}\label{eq.x.sr.dt}
	x^i(t+\diff t) =x^i(t)+
	\diff t \cdot L^{ij} E_{x^j}
	+\frac{\tau\cdot \diff t}{2} \frac{\delta}{\delta x^k}\left(L^{ij}E_{x^j}\right)L^{kl}E_{x^l},
\end{equation}
which resembles the formal solution of the original reversible equation \eqref{eq.xtau}.
\begin{itemize}
	\item Indeed, if $\diff t= \tau$, both equations are the same. Recall that numerical scheme \eqref{eq.xtau} conserves both energy and entropy up to the order $\Obig(\tau^2)$ while the regularized evolution equation \eqref{eq.x.evo.sr} with or without ``internal entropy'' (see the following Section, Eq. \eqref{eq.s.evo.sre}) either reduce energy and produce entropy or keep the energy and raise entropy.  This case is actually just a prolongation of the Hamiltonian vector field leading to a second-order reversible scheme.
	\item Therefore, taking $\diff t< \tau$, scheme \eqref{eq.x.sr.dt} produces entropy and either reduces energy or conserves energy (in the extended sense as in Sec. \ref{sec.sin}). This choice of $\tau$ can be interpreted as smoothing the evolution over a period of time, and an emergence of overall behavior can thus be anticipated.
	\item Taking $\diff t> \tau$, the scheme reduces entropy and produces or conserves energy. One should therefore always take $\diff t \leq \tau$. The choice of $\tau<\diff t$ is against the notion of smoothing, discretization step is coarser than the time-scale for smoothing of trajectories, and does not indeed make much sense and should not be used.
\end{itemize}

\subsection{Ehrenfest regularization with internal entropy}\label{sec.sin}
Under the above assumptions (energy be a convex functional and entropy be a concave Casimir functional), entropy is produced while energy is reduced. In order to recover the physical constraint of energy conservation, it is necessary to include internal entropy (density) $s_{in}$ among the state variables. Total energy $E^{(tot)}(\xx,s_{in})$ then depends besides the original state variables also on the internal entropy density. Evolution equation for the internal entropy density is purely dissipative
\begin{equation}\label{eq.s.evo.sre}
	\dot{s}_{in} = \frac{1}{E^{(tot)}_{s_{in}}} \frac{\tau}{2} \left(\LL\cdot E_\xx\right) \cdot \frac{\delta^2 E}{\delta \xx \delta \xx}\cdot \left(\LL\cdot E_\xx\right)
\end{equation}
so that conservation of total energy $E^{(tot)}(\xx,s_{in}) = E(\xx)+E_{in}(s_{in})$
\begin{equation}
	\dot{E}^{(tot)} = \dot{E}(\xx) + E^{(tot)}_{s_{in}} \dot{s}_{in} = 0
\end{equation}
is fulfilled. This approach to entropy production is successfully used for instance in the SHTC framework \cite{ADER,CMAT2018}. Evolution of total entropy $S^{(tot)}(\xx,s_{in})=S+s_{in}$ is then
\begin{equation}
	\dot{S}^{(tot)} = \dot{S}(\xx) + \dot{s}_{in} \geq 0.
\end{equation}
By including the internal entropy density $s_{in}$ among the state variables, we recover the energy conservation law while keeping the second law of thermodynamics. Actually, the second law is obtained this way also for constant Poisson bivectors. This \textbf{Ehrenfest-regularized evolution with internal entropy}, which consists of reversible Hamiltonian evolution and irreversible evolution, conserves energy and produces entropy.

The regularization with internal entropy is discretized as
\begin{subequations}\label{eq.sr.rn.dt}
\begin{eqnarray}
	x^i(t+\diff t) &=&x^i(t)+
	\diff t \cdot L^{ij} E_{x^j}
	+\frac{\tau\cdot \diff t}{2} \frac{\delta}{\delta x^k}\left(L^{ij}E_{x^j}\right)L^{kl}E_{x^l}\Big|_{\xx(t)}\\
	s_{in}(t+\diff t)  &=&s_{in}(t)+ \frac{\diff t}{E^{(tot)}_{s_{in}}} \frac{\tau-\diff t}{2} \left(\LL\cdot E_\xx\right) \cdot \frac{\delta^2 E}{\delta \xx \delta \xx}\cdot \left(\LL\cdot E_\xx\right)\Big|_{\xx(t)}.
\end{eqnarray}
\end{subequations}
The last term in the latter equation is proportional to the $(\tau-\diff t)$ difference so that the scheme becomes reversible with the choice $\tau=\diff t$ as desired (see the previous section).

\subsection{Energetic Ehrenfest regularization}
The regularized evolution equations \eqref{eq.x.evo.sr} can also be split as
\begin{eqnarray}\label{eq.EhRe.split}
	\dot{x}^i &=& L^{ij}E_{x^j} - \frac{\tau}{2}\underbrace{L^{ji} E_{x^j x^k} L^{kl}}_{\stackrel{def}{=}M^{il}} E_{x^l}
	+\frac{\tau}{2} \underbrace{\frac{\delta L^{ij}}{\delta x^k} L^{kl}E_{x^l}}_{\stackrel{def}{=}N^{ij}} E_{x^j}\nonumber\\
	&=&L^{ij}E_{x^j} - \frac{\tau}{2} M^{ij}E_{x^j} + \frac{\tau}{2} N^{ij} E_{x^j},
\end{eqnarray}
where a symmetric operator $M^{ij}=M^{ji}$ and an antisymmetric operator $N^{ij}=-N^{ji}$ were identified. Let us now discuss properties of the operators.

\subsubsection{Evolution equations}
Assuming that energy is a convex functional of the state variables, the $\MM-$operator,
\begin{equation}\label{eq.Mij}
	M^{ij} = L^{ki} E_{x^k x^l} L^{lj},
\end{equation}
is positive definite for
\begin{equation}
	v_i M^{ij} v_j = L^{ki}v_i E_{x^k x^l} L^{lj}v_j \geq 0\qquad \forall v_i.
\end{equation}
Therefore, the operator causes the reduction of energy in the regularized evolution, since
\begin{equation}
	\dot{E} = -\frac{\tau}{2}E_{x^i} M^{ij} E_{x^j} \leq 0.
\end{equation}
Note that neither the $\LL-$operator nor the $\NN-$operator contributes to the change of total energy due to their antisymmetry. Concerning entropy, which is a Casimir of the Poisson bracket, the $\MM$-operator does not affect it, since
\begin{equation}
	S_{x^i} M^{ij} E_{x^j} = \underbrace{S_{x^i}L^{ji}}_{=0} E_{x^l x^k} L^{kl}E_{x^l} = 0.
\end{equation}
The evolution equations with only the $\MM$-operator, let us call them the \textbf{energetic Ehrenfest regularization} (E-EhRe), are
\begin{equation}\label{eq.EhRe.M}
	\dot{x}^i = L^{ij}E_{x^j} - \frac{\tau}{2}M^{ij}E_{x^j},
\end{equation}
and they conserve Casimirs while reducing (convex) energy, $\dot{S}=0$, $\dot{E}\leq 0$.
The $\MM$-operator, which is symmetric and positive definite, reduces total energy while keeping (kinetic) entropy constant.

\subsubsection{Numerical scheme}
Let us now discuss possible discretizations of the energetic Ehrenfest-regularized evolution equations \eqref{eq.EhRe.M}. It can be shown by straightforward calculation, by repeating calculations in \eqref{Eq.18} and \eqref{Eq.19}, that the forward (explicit) Euler discretization with $\diff t<\tau$ reduces energy, $\diff t=\tau$ does not change energy and $\diff t > \tau$ raises energy. Moreover, the forward Euler scheme reduces entropy regardless the choice of $\tau$. Backward (implicit) Euler scheme reduces energy while raising entropy. Let us consider the following ``semi-implicit Crank-Nicolson scheme'' (forward Euler in $\diff t$, implicit in $\tau$)
\begin{equation}\label{eq.EhRe.M.CN}
	x^i(t+\diff t)= x^i(t) + \frac{\diff t}{2} \left( L^{ij} E_{x^j} - \frac{\tau}{2}M^{ij} E_{x^j}\right)\Big|_{\xx(t)}
	+ \frac{\diff t}{2} \left( L^{ij} E_{x^j} - \frac{\tau}{2}M^{ij} E_{x^j}\right)\Big|_{\xx(t+\diff t)}.
\end{equation}
It is a matter of straightforward calculation to show that
\begin{subequations}
	\begin{eqnarray}
		E(\xx(t+\diff t))&=&E(\xx(t))\underbrace{-\frac{\diff t \cdot \tau}{2} L^{ij}E_{x^j} E_{x^i x^k} L^{kl}E_{x^l}}_{\leq 0} + \Obig(\tau)^3,\\
		S(\xx(t+\diff t))&=&S(\xx(t)) + \Obig(\tau)^3.
	\end{eqnarray}
\end{subequations}
The Crank-Nicolson scheme is thus a suitable choice for the E-EhRe evolution as it conserves the properties of E-EhRe evolution to its discretization (keeps entropy while reducing energy).

\subsubsection{E-EhRe with internal entropy}
The energetic Ehrenfest regularization reduces the energy of the system. More precisely, it usually reduces the kinetic energy, which serves as a generator of the Hamiltonian dynamics. Since the total energy must be conserved, one can introduce an internal entropy $s_{in}$ that is advected by the Hamiltonian dynamics and grows so that the total energy $E_{tot} = E + E_{in}$ is conserved,
\begin{subequations}\label{eq.EEhRe.s}
	\begin{eqnarray}
		\dot{x}^i &=& L^{ij} E_{x^j} - \frac{\tau}{2}M^{ij} E_{x^j},\\
		\dot{s}_{in} &=& \frac{1}{\frac{\delta E_{in}}{\delta s_{in}}} E_{x^i} \frac{\tau}{2}M^{ij} E_{x^j}.
	\end{eqnarray}
\end{subequations}
Additional terms in evolution equations can appear but cannot contribute both to energy and to entropy balance. Such ``reversible'' terms can be those expressing passive advection of entropy by $\xx$ (hence not contributing to total entropy).
 For instance, it is zero in the case of rigid body motion while being equal to $-\partial_j( s_{in} v_j)$ in the case of fluid mechanics. Total energy is conserved by construction while entropy is produced.

\subsubsection{Stability by the Energy-Casimir method}
Energy is reduced in the energetic regularization while entropy 
is kept constant. Assume that the potential $\Phi$ in the Energy-Casimir method is constructed as $E-\alpha S$. The potential then decreases in time.

If it is moreover concave in a neighborhood of the stationary state $\xx_0$, then, because it further decreases, it becomes a less and less restrictive bound on the distance from the stationary point. The Energy-Casimir method then does not say anything about the stability of the stationary point (or it indicates instability).

If, on the other hand, the potential is convex, the evolution not only stays in the neighborhood of the stationary point, but it also approaches to a point where $\dot{\Phi} = 0$ because of the decrease of $\Phi$. The stationary point $\xx_0$ is thus stable and nearly asymptotically stable.

\subsubsection{Relation to GENERIC}
The energetic regularization with internal energy \eqref{eq.EEhRe.s} can also be seen as a particular realization of the GENERIC framework, see \cite{GO,OG,hco,PKG} or Sec. \ref{sec.discussion}. Indeed, after the transformation to the entropic representation (with total energy density as a state variable), the irreversible terms can be seen as derivatives of a quadratic dissipation potential with respect to conjugate variables (derivatives of entropy), see \cite{CMAT2018}, multiplied by a temperature prefactor.

\subsection{Entropic Ehrenfest regularization}
Let us now analyze the other part of the split EhRe evolution \eqref{eq.EhRe.split}, which involves the antisymmetric $\NN-$operator.

\subsubsection{Evolution equations}
The regularized evolution equations involving only Hamiltonian and the $\NN-$operator parts are
\begin{equation}\label{eq.EhRe.N}
	\dot{x}^i = L^{ij} E_{x^j} + \frac{\tau}{2}N^{ij} E_{x^j}.
\end{equation}
We shall now discuss properties of these equations, which are referred to as the \textbf{Entropic Ehrenfest regularization} (S-EhRe).

First, the S-EhRe equations raise entropy, since
\begin{equation}
	\dot{S} = \frac{\tau}{2}S_{x^i} N^{ij} E_{x^j} = \frac{\tau}{2}\underbrace{\frac{\delta}{\delta x^k}\left(S_{x^i} L^{ij}\right)}_{=0} L^{kl} E_{x^l} E_{x^j} - \frac{\tau}{2}S_{x^i x^k} L^{ij} E_{x^j} L^{kl} E_{x^l} \geq 0
\end{equation}
due to the assumed concavity of entropy (and negative definiteness of its second differential). Secondly, energy is conserved by the Hamiltonian part as well as by the irreversible part due to the antisymmetry with respect to swapping $i$ and $j$. Assuming that (kinetic) entropy is a concave Casimir of the Poisson bracket, the entropic regularized evolution equations \eqref{eq.EhRe.N} conserve energy while producing (kinetic) entropy.

Note that the $\NN-$operator  can also be rewritten as the Poisson bracket of the Poisson bivector and energy,
\begin{equation}\label{eq.Nij}
	N^{ij} = \frac{\delta L^{ij}}{\delta x^k} L^{kl} E_{x^l} = \{L^{ij},E\},
\end{equation}
which simplifies the explicit calculation of the operator.

\subsubsection{Numerical scheme}
Let us now discuss possible discretizations of the  regularized evolution equations \eqref{eq.EhRe.N}. It can be shown that forward (explicit) Euler discretization with $\tau=\diff t$ leads to a scheme that raises energy while keeping constant entropy up to the order $\Obig(\tau^2)$. Backward (implicit) Euler scheme, on the other hand, reduces energy while raising entropy. The desired properties has again the semi-implicit Crank-Nicolson scheme
\begin{equation}\label{eq.EhRe.N.CN}
	x^i(t+\diff t)= x^i(t) + \frac{\diff t}{2} \left( L^{ij} E_{x^j} + \frac{\tau}{2}N^{ij} E_{x^j}\right)\Big|_{\xx(t)}
	+ \frac{\diff t}{2} \left( L^{ij} E_{x^j} + \frac{\tau}{2}N^{ij} E_{x^j}\right)\Big|_{\xx(t+\diff t)}.
\end{equation}
It is a matter of straightforward calculation to show that
\begin{subequations}
	\begin{eqnarray}
		E(\xx(t+\tau))&=&E(\xx(t))+\Obig(\tau)^3\\
		S(\xx(t+\tau))&=&S(\xx(t)) \underbrace{-\frac{\tau^2}{2}S_{x^i x^k} L^{ij}E_{x^j} L^{kl}E_{x^l}|_{\xx(t)}}_{\geq 0} + \Obig(\tau)^3
	\end{eqnarray}
\end{subequations}
For the choice $\tau = \diff t$ scheme \eqref{eq.EhRe.N.CN} thus conserves energy and produces entropy.

\subsubsection{Stability by Energy-Casimir method}
Entropy (a concave Casimir of the Poisson bracket) grows in the entropic Ehrenfest regularization. Assume that the potential $\Phi$ in the Energy-Casimir method is constructed as $E-\alpha S$, $\alpha < 0$. The potential then grows as entropy grows.

If it is moreover concave in a neighborhood of the stationary state $\xx_0$, it grows until it reaches a state in the neighborhood where $\dot{\Phi} =0$. The stationary point is thus stable and nearly asymptotically stable.

If, on the other hand, the potential is convex, it grows and the evolution may eventually leave the neighborhood of the stationary point. The Energy-Casimir method then does not tell anything about the stability of the stationary point (or it indicates instability).

\subsubsection{Relation to GENERIC}
Also the entropic Ehrenfest regularization can be seen as a realization of the GENERIC framework when the underlying Poisson bracket is Lie-Poisson. This is demonstrated on the example of rigid body motion in Sec. \ref{sec.compare}. Entropy, a Casimir of the bracket, then comes out of derivatives of the Poisson bivector.

\section{Classical mechanics}
Let us now demonstrate the Ehrenfest-regularized Hamiltonian evolution on classical mechanics.

\subsection{Hamilton canonical equations}
Motion of a particle in classical mechanics is described by Hamilton canonical equations,
\begin{subequations}
	\begin{eqnarray}
		\dot{q} &=& E_{p}\\
		\dot{p} &=& -E_{q},
	\end{eqnarray}
	where $E$ is the energy of that particle.
\end{subequations}
Hamilton canonical equations can be seen as generated by Poisson bivector
\begin{equation}
	\LL =
	\begin{pmatrix}
		0 & 1\\
		-1 & 0
	\end{pmatrix}
\end{equation}
or the canonical Poisson bracket
\begin{equation}
	\{F,G\} = F_q \cdot G_p - G_q \cdot F_p.
\end{equation}
The bivector is a regular matrix, which means that there are no Casimirs of the Poisson bracket (no place for kinetic entropy). Therefore, we choose the energetic Ehrenfest regularization (E-EhRe).

\subsection{Energetic Ehrenfest regularization}
The $\MM-$operator is constructed as
\begin{equation}
	\MM = \LL^T \cdot \diff^2 E \cdot \LL.
\end{equation}
Choosing energy (a particular physical system) and relaxation time then makes it possible to write down the E-EhRe evolution equations explicitly.

\subsubsection{Oscillator with damping}
Energy of a particle in a potential field is given by
\begin{equation}
	E = \frac{p^2}{2m} + V(q).
\end{equation}
Assuming that $V(q)$ be convex\footnote{Non-convex energy is commented in Sec. \ref{sec.nonconvex}.}, the particle undergoes oscillatory motion, therefore being an oscillator. It is expected that due to some friction (interaction with surroundings) the particle will gradually slow down until it ends up in the minimum of the potential field.

The second differential of energy is
\begin{equation}
	\diff^2 E =
	\begin{pmatrix}
		V_{qq} & 0 \\
		0 & \frac{1}{m},
	\end{pmatrix}
\end{equation}
and the $\MM-$operator is then
\begin{equation}
	\MM =
	\begin{pmatrix}
		\frac{1}{m}&0\\
		0 & V_{qq}
	\end{pmatrix}.
\end{equation}
Finally, the E-EhRe evolution equations for the oscillator become
\begin{subequations}\label{eq.EhRe.qp}
	\begin{eqnarray}
		\dot{q} &=& E_{p} - \frac{\tau}{2m} V_q\\
		\dot{p} &=& -E_{q} -\frac{\tau}{2m} V_{qq} p
	\end{eqnarray}
\end{subequations}

Energy is clearly dissipated by the equations as
\begin{equation}
	\dot{E} = -\frac{\tau}{2} \left(\frac{(V_q)^2}{m} + V_{qq}\left(\frac{p}{m}\right)^2\right).
\end{equation}
Therefore, we can add internal entropy as follows
\begin{subequations}\label{eq.can.EEhRe.s}
	\begin{eqnarray}
		\dot{q} &=& E_{p} - \frac{\tau}{2m} V_q\\
		\dot{p} &=& -E_{q} -\frac{\tau}{2m} V_{qq} p\\
		\dot{s}_{in} &=& \frac{1}{E^{(tot)}_{s_{in}}} \frac{\tau}{2} \left(\frac{(V_q)^2}{m} + V_{qq}\left(\frac{p}{m}\right)^2\right),
	\end{eqnarray}
	where the $E^{(tot)}(q,p,s_{in})=p^2/2m + V(q) + E_{in}(s_{in})$ stands for the total energy of the oscillator.
\end{subequations}

Note that these equations should be solved using the Crank-Nicolson scheme so that properties (reduction of kinetic energy) of the equations are satisfied also by the numerical scheme.

Equations \eqref{eq.can.EEhRe.s} clearly drive the evolution towards the state with lowest kinetic energy, $p=0$ and $q$ where $V$ has the minimum, which is the expected physical behavior.

\section{Rigid body motion}
Motion of a freely rotating body is usually described by evolution of Euler angles (kinematics) and evolution of angular momentum of the body (dynamics), see e.g. \cite{landau1}. Since, however, kinetic energy depends only on the angular momentum (or angular velocity), the evolution equation for angular momentum forms a system of closed equations itself. Let us thus for simplicity focus only dynamics of angular momentum of the body.

\subsection{Reversible evolution}
\subsubsection{Lie-Poisson bracket}
Consider a rotating rigid body. Evolution of the angular momentum vector $\mm$ regarded from the coordinate system attached to the body is expressed by Lie-Poisson bracket (see \cite{arnold})
\begin{equation}\label{eq.PB}
\{F,G\}^{(SO(3))}=
- m_i \eps^{ijk} \frac{\partial F}{\partial m_j}\frac{\partial G}{\partial m_k},
\end{equation}
where $F(\mm)$ and $G(\mm)$ are two arbitrary smooth functions of $\mm$, and where $\eps^{ijk}$ is the Levi-Civita symbol. See \ref{sec.Ham.SO3} for details. The Poisson bivector corresponding to this bracket is equal to
\begin{equation}\label{eq.m.L}
	L^{ij} = -m_k \eps^{kij}.
\end{equation}
Evolution of an arbitrary function $F(\mm)$ is then generated by
\begin{equation}
\dot{F} = \{F,E\}^{(SO(3))},
\end{equation}
where $E(\mm)$ is the energy of the rotating body. Evolution of angular momentum regarded from the coordinate system attached to the body is Hamiltonian.

\subsubsection{Reversible evolution equations}
Evolution of an arbitrary function $F$ can be expressed also as
\begin{equation}
\dot{F} = \frac{\partial F}{\partial m_j} \dot{m}_j,
\end{equation}
and by comparing with Eq. \eqref{eq.PB} we obtain the reversible evolution equation of $\mm$
\begin{equation}\label{eq.m.evo}
\dot{m}_j = -\eps_{ijk} m_i E_{m_k}, \qquad \mbox{or} \qquad \dot{\mm} = \mm\times E_{\mm}.
\end{equation}
Since the tensor of inertia is diagonal in the frame attached to the body, the kinetic energy (or Hamiltonian) is
\begin{equation}\label{eq.SO3.E}
E = \frac{1}{2}\left(\frac{m_x^2}{I_x}+\frac{m_y^2}{I_y}+\frac{m_z^2}{I_z}\right), \quad I_x\leq I_y \leq I_z,
\end{equation}
where $I_x$, $I_y$ and $I_z$ represent the main moments of inertia of the body. In particular, derivatives of energy with respect to $\mm$ are the angular velocities of rotation around the three axes,
\begin{equation}
\omega_i = \frac{\partial E}{\partial m_i} = \frac{m_i}{I_i},
\end{equation}
and the evolution equation for angular momentum $\mm$ can be then rewritten as
\begin{equation}
\dot{\mm} = \mm\times \omega.
\end{equation}

\subsubsection{Conservation laws}
Consider a function of the magnitude of $\mm$, $S(\mm^2)$. Evolution of such a function is then
\begin{equation}
	\dot{S} = \{S,E\}^{(SO(3))} = -m_i \eps_{ijk} \frac{\partial S}{\partial (\mm^2)} 2 m_j E_{m_k} = 0,
\end{equation}
which is zero for any choice of energy due to the 
antisymmetry with respect to swapping $i\leftrightarrow j$.
 Bracket \eqref{eq.PB} has thus Casimir functions $S(\mm^2)$, which are conserved regardless the choice of energy. This means that the angular momentum seen from the body does not change its magnitude although it can change its direction.

Moreover, energy is automatically conserved due to the antisymmetry of the underlying Lie-Poisson bracket,
\begin{equation}
\dot{E} = \{E,E\}^{(SO(3))} = -\{E,E\}^{(SO(3))} = 0.
\end{equation}

A system of isolated particles, as for instance the free rigid body, is known to conserve its angular momentum \cite{landau1} regarded from an inertial frame (not from the frame attached to the body). How is the angular momentum conservation realized in the Hamiltonian rigid body dynamics \eqref{eq.m.evo}?

Firstly the magnitude of angular momentum $\mm^2$ (as well as any function of it) is conserved regardless the choice of energy.
Secondly, in order to talk also about direction of the angular momentum regarded from an inertial reference frame, we first need to introduce at least two position vectors $\rr_a$ and $\rr_b$ in the inertial frame into the state variables. These two position vectors are then sufficient to express the angle of rotation of the solid body. The Poisson bracket has to be also extended so that it expresses also evolution of the position vectors,
\begin{eqnarray}
	\{F, G\}^{(\mbox{heavy top})} &=& -\mm \cdot \left(F_\mm \times G_\mm\right)\nonumber\\
	&&-\rr_a \cdot \left(F_\mm \times G_{\rr_a}-G_\mm \times F_{\rr_a}\right)\nonumber\\
	&&-\rr_b \cdot \left(F_\mm \times G_{\rr_b}-G_\mm \times F_{\rr_b}\right),
\end{eqnarray}
which is the heavy top Poisson bracket, see e.g. \cite{Holm-stability}. Casimirs of this bracket are not only functions of $\mm^2$, but also of $\rr_a\cdot\mm$ and $\rr_b\cdot\mm$, which means that not only magnitude of angular momentum, but also its projection to the two position vectors is constant. In other words, the angular momentum regarded from an inertial frame is constant although, of course, it varies when regarded from the frame attached to the body. Magnitude of the angular momentum is constant in both frames.

\subsubsection{Stability by Energy-Casimir method}\label{sec.SO3.stab}
Stability of Hamiltonian evolution \eqref{eq.m.evo} can be proved for instance by the Energy-Casimir method \cite{Holm-stability}, which shows that rotation around the axis with lowest ($I_x$) and highest ($I_z$) inertia are stable while rotation around the axis with the middle inertia ($I_y$) is unstable.

However, assuming the same magnitudes of $\mm$, body rotating around the axis with lowest moment of inertia has higher energy than rotation around the axis with highest moment of inertia. Therefore, it can be anticipated that only rotation around the axis with highest inertia will be stable in reality. This has been observed for instance in the textbook \cite{landau5} or in the unexpected change of rotation of the Explorer 1 probe, see e.g. \cite{Krechetnikov-Marsden}, \cite{Bloch-dissipation} or \cite{Efroimsky}.

The loss of stability of rotation around the axis with lowest moment of inertia is connected with growth of some entropy, i.e. the second law of thermodynamics. But the Lie-Poisson dynamics \eqref{eq.m.evo} is reversible and does not produce any entropy. The solution is thus to equip the original Hamiltonian dynamics with dissipative terms, as was done for instance in \cite{Materassi}. The question is, however, how to add the dissipation. In this paper we suggest that the Ehrenfest regularization (in particular the E-EhRe equations \eqref{eq.EhRe.M}) is an appropriate way.

Before going to the stability analysis itself, let us first discuss sufficient conditions for entropy to be concave. Assuming that the form of entropy is
\begin{equation}
	S(\mm) = \eta(\mm^2),
\end{equation}
with $\eta$ being a function, then second differential of entropy becomes
\begin{equation}\label{eq.d2S}
	\frac{\partial^2 S}{\partial m_i \partial m_k} = 2\left(\eta' \delta_{ik} + 2 m_i m_k \eta''\right).
\end{equation}
A necessary condition for this matrix to be negative-semidefinite is that $\eta'\leq 0$, since otherwise a vector perpendicular to $\mm$ would lead to $\vv\cdot\diff^2 S\cdot \vv \geq 0$. A sufficient condition for that matrix to be negative-semidefinite is that $\eta'\leq 0$ and $\eta''\leq 0$.

Let us now reproduce the algorithm of the Energy-Casimir method in the case of reversible rigid body rotation (shown in \cite{Holm-stability}). Assume that entropy is a strictly concave Casimir $S(\mm) = \eta(\mm^2)$. The potential in the Energy-Casimir method then reads
\begin{equation}
	\Phi(\mm) = -\alpha \eta(\mm^2) + \sum_i \frac{1}{2}\frac{m^2_i}{I_i}.
\end{equation}
The stationary points are pure rotations around the three axes. For $\xx_0 = (0, 0, m^0_z)$, gradient of $\Phi$,
\begin{equation}
	\begin{pmatrix}
		m_x\left(-2\alpha\eta' +  \frac{1}{I_x}\right)\\
		m_y\left(-2\alpha\eta' +  \frac{1}{I_y}\right)\\
		m_z\left(-2\alpha\eta' +  \frac{1}{I_z}\right)
	\end{pmatrix}
	=
	\begin{pmatrix}
		0\\
		0\\
		0
	\end{pmatrix},
\end{equation}
vanishes at $\xx_0$ if $\eta' = \frac{1}{2\alpha I_z}$. Therefore, the Casimir is chosen as
\begin{equation}
	\eta(\mm^2) = -\frac{1}{2 I_z} \mm^2 + \frac{\beta}{2} ((\mm)^2-(\xx_0)^2)^2,
\end{equation}
$0< \beta < \frac{1}{2I_z}$, $\alpha = -1$, so that the entropy is concave. The second differential of the potential (using \eqref{eq.d2S}),
\begin{equation}
	\diff^2 \Phi =
	\begin{pmatrix}
		\frac{1}{I_x} - \frac{1}{I_z} & 0 & 0\\
		0 & \frac{1}{I_y} - \frac{1}{I_z}  & 0\\
		0 & 0 & 2\beta (m_z^0)^2
	\end{pmatrix}
\end{equation}
is then positive definite (strictly),
since $I_x\leq I_y\leq I_z$. Although being difference of two convex function, the potential $\Phi$ is convex. The quadratic form $Q$ can be chosen as equal to $\Phi$, which is also quadratic, strictly convex and equal to zero at $\xx_0$. The point (rotation around the major axis) is thus Lyapunov stable.

In the case $\xx_0 = (m_x, 0, 0)$, the Casimir is chosen as
\begin{equation}
	\eta(\mm^2) = -\frac{1}{2I_x}\mm^2 - \frac{\beta}{2}((\mm)^2-(\xx_0)^2)^2,
\end{equation}
$\beta > 0$, and $\alpha = -1$. The second differential of the potential $\Phi$ is then negative definite (strictly). Negative of the potential, $-\Phi$, plays the role of a Lyapunov functional, $Q=-\Phi$, and rotation around the minor axis is thus Lyapunov stable.

Finally, second differential of $\Phi$ constructed for $\xx_0 = (0, m_y, 0)$ is indefinite and the Energy-Casimir method does not tell anything about stability of that state (or indicates instability, which can be proved using eigenvalue analysis of the linearized equations).

Note that the value of $\beta$ in the numerical simulations below is taken as $I_z/4$.

In summary, the Energy-Casimir method proves non-linear stability of rotation around the major ($z$) and minor ($x$) axes while telling nothing about (or suggesting possible instability of) the rotation around the middle axis ($y$).

\subsubsection{Instability by eigenvalues}\label{sec.SO3.eig}
In the preceding section we recalled the Energy-Casimir method, which is a tool for proving nonlinear stability. In this section we recall the standard results on linearized stability, which, in contrast to the Energy-Casimir method, can prove instability of critical points.

Taking $\mm^{0} = (0,0,m^0_z)$, equations \eqref{eq.m.evo} linearized by $\mm = \mm^0 + \delta\mm$ become
\begin{equation}
	\frac{\diff}{\diff t}
	\begin{pmatrix}
		\delta m_x\\
		\delta m_y\\
		\delta m_z
	\end{pmatrix}
	=
	m_z^0 \begin{pmatrix}
		0 & J_x  & 0\\
		J_y & 0 & 0\\
		0 & 0 & 0
	\end{pmatrix}
	\cdot
	\begin{pmatrix}
		\delta m_x\\
		\delta m_y\\
		\delta m_z
	\end{pmatrix},
\end{equation}
where $J_x = \frac{1}{I_z}-\frac{1}{I_y}$ and $J_y$ and $J_z$ are obtained by cyclic permutation. From $I_x<I_y<I_z$ we obtain that $J_x<0$, $J_y > 0$ and $J_z < 0$. The nontrivial (there is always one zero eigenvalue) eigenvalues of the matrix are given by
\begin{equation}
	\lambda^2 = J_x J_y < 0,
\end{equation}
which means that they are purely imaginary. There is no unstable mode.

Taking $\mm^0 = (m^0_x, 0, 0)$, eigenvalues of the corresponding matrix fulfill
\begin{equation}
	\lambda^2 = J_y J_z < 0,
\end{equation}
which means that they are also purely imaginary. There is no unstable mode either.

Taking $\mm^0 = (0, m^0_y, 0)$, the nontrivial eigenvalues fulfill
\begin{equation}
	\lambda^2 = J_z J_x > 0,
\end{equation}
which means that there is one positive eigenvalue and one negative. The former is an unstable mode, which means that rotation around the middle axis is unstable.

In summary, the Energy-Casimir method shows stability of rotation around the major and minor axes ($z$ and $x$) while eigenvalues of the linearized equations show instability or rotation around the middle axis ($y$).

\subsection{Irreversible evolution by Ehrenfest regularization}

\subsubsection{Hamiltonian Ehrenfest regularization}
Ehrenfest regularization of Hamiltonian evolution equation \eqref{eq.m.evo} is given by plugging the Poisson bivector \eqref{eq.m.L} and energy \eqref{eq.SO3.E} into Eq. \eqref{eq.x.evo.sr}. For the $m_x$ component of the angular momentum it becomes
\begin{equation}\label{eq.m.sr}
	\dot{m}_x = m_y m_z J_x + \frac{\tau}{2} m_x J_x J_y J_z\left(\frac{m^2_y}{J_y} + \frac{m^2_z}{J_z}\right),
\end{equation}
where $J_x = \frac{1}{I_z}-\frac{1}{I_y}$ and $J_y$ and $J_z$ are obtained by cyclic permutation. Evolution equations for $m_y$ and $m_z$ are also obtained by cyclic permutations of Eq. \eqref{eq.m.sr}.

Since kinetic energy is dissipated by equations \eqref{eq.m.sr}, internal entropy can be added,
\begin{equation}\label{eq.SO3.SR.s}
	\dot{s}_{in} = \frac{1}{E^{(tot)}_{s_{in}}} \frac{\tau}{2}\left(\frac{(m_y m_z J_x)^2}{I_x}+\frac{(m_z m_x J_y)^2}{I_y}+\frac{(m_x m_y J_z)^2}{I_z}\right),
\end{equation}
so that total energy $E^{(tot)} = E + E_{in}(s_{in})$ is conserved. Assuming positive temperature, $\frac{\partial E_{in}}{\partial s_{in}}> 0$, internal entropy clearly grows. When the regularized equations \eqref{eq.m.sr} are equipped with the evolution equation for internal entropy, Eq. \eqref{eq.SO3.SR.s}, the total energy is conserved while internal entropy being produced.


\subsubsection{Numerical solution}\label{sec.SO3.EhRe.num}

The regularized (EhRe) evolution equation \eqref{eq.m.sr} is now to be solved numerically. Forward Euler discretization is a particular case of Eqs. \eqref{eq.sr.rn.dt},
\begin{subequations}
\begin{eqnarray}
	\label{eq.mx.tau} m_x(t+\diff t) &=& m_x(t) + \diff t \cdot m_y(t) m_z(t) J_x \nonumber\\
	&&+ \frac{\diff t \cdot \tau}{2} m_x(t) J_x J_y J_z\left(\frac{m^2_y(t)}{J_y} + \frac{m^2_z(t)}{J_z}\right)\\
	s_{in}(t+\diff t) &=& s_{in}(t)+ \nonumber\\
	&&\diff t\cdot \frac{\tau-\diff t}{2 E^{(tot)}_{s_in}(t)} \left(\frac{(m_y m_z J_x)^2}{I_x}+\frac{(m_z m_x J_y)^2}{I_y}+\frac{(m_x m_y J_z)^2}{I_z}\right).
\end{eqnarray}
\end{subequations}
Numerical results stemming from this scheme are presented below.

\begin{figure}[ht]
	\centering
	\begin{subfigure}{.45\textwidth}
		\centering
		\includegraphics[width=0.95\linewidth]{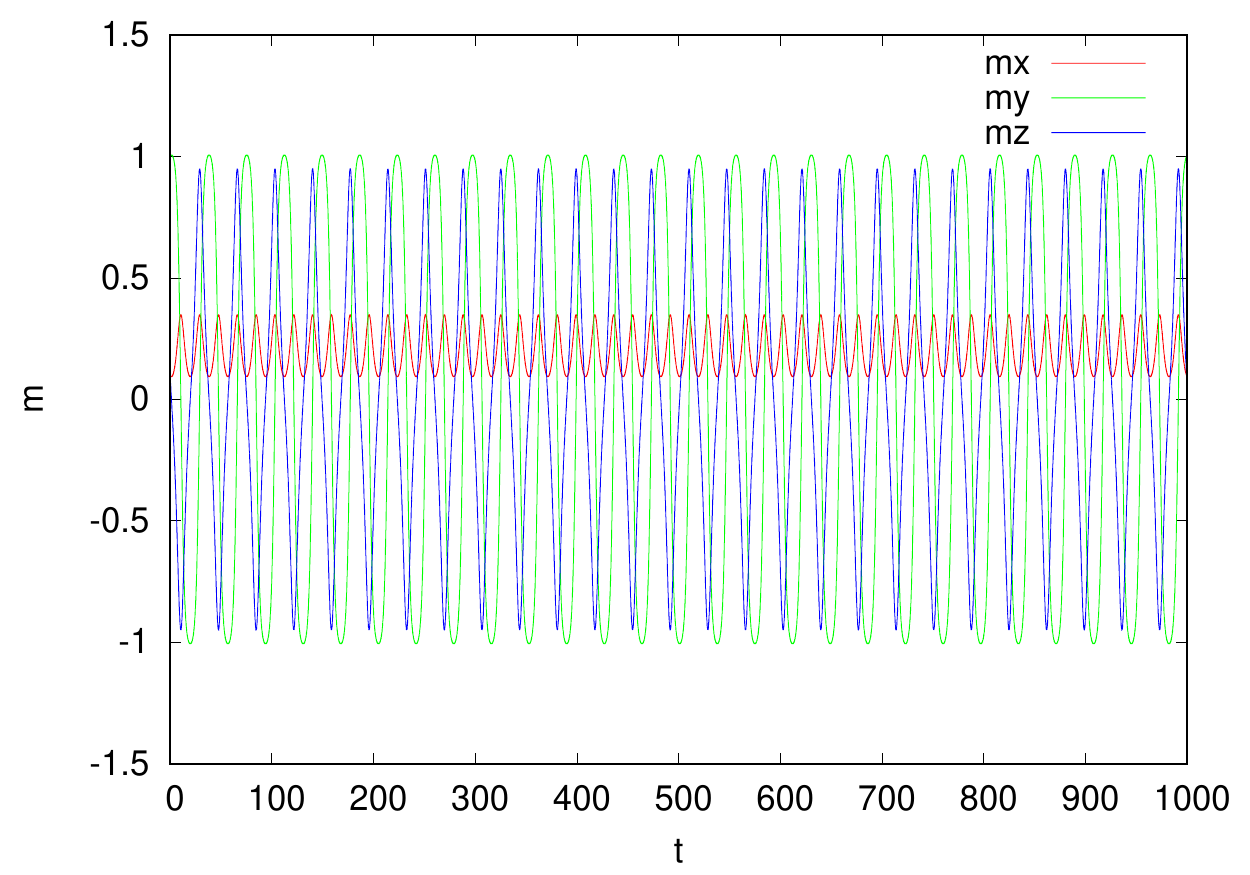}
		\caption{\label{fig.m.1dt}
		Components ($m_x$ red, $m_y$ green and $m_z$ blue) of angular momentum in time. The solution is periodic.}
	\end{subfigure}%
\hspace{5pt}\begin{subfigure}{.45\textwidth}
		\centering
		\includegraphics[width=0.95\linewidth]{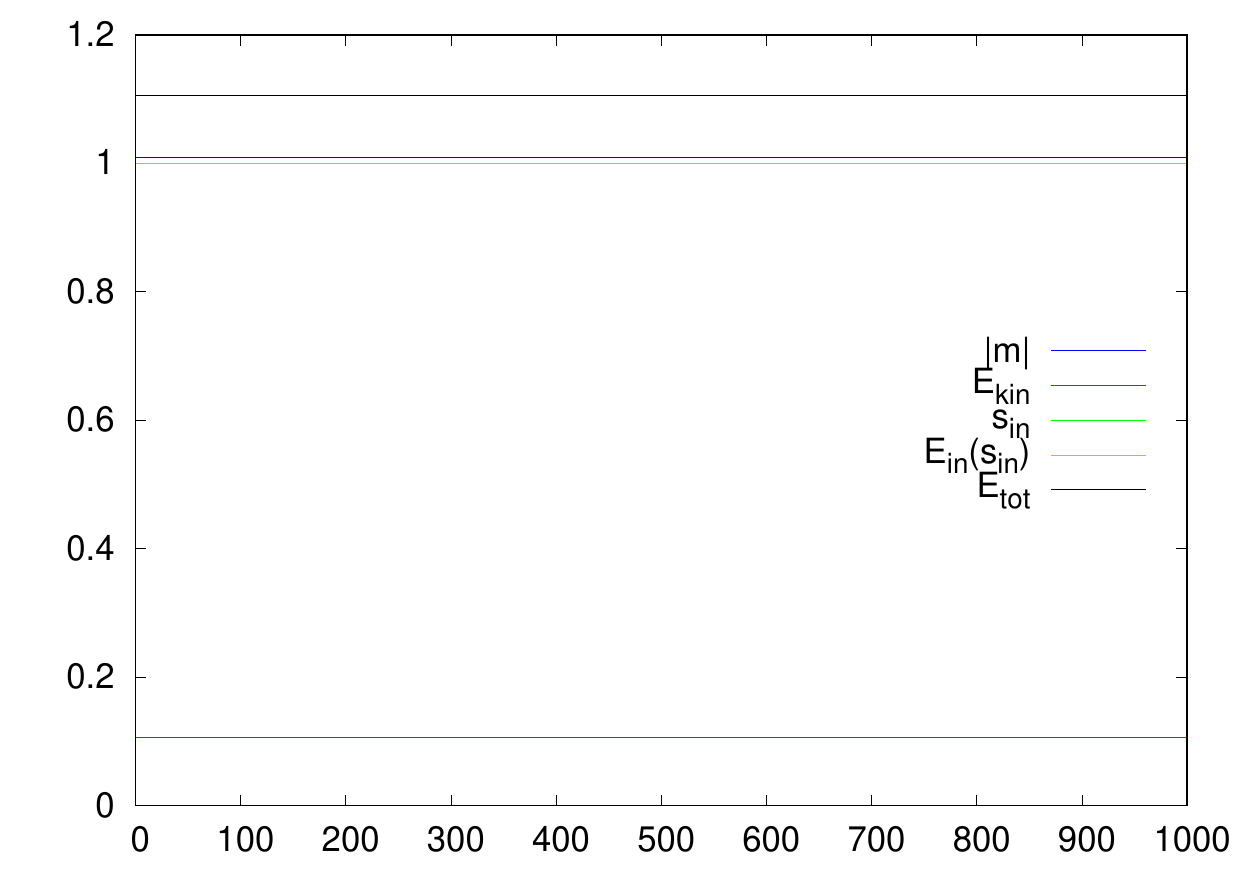}
		\caption{\label{fig.Em.1dt}
		Kinetic energy (red) and magnitude of angular momentum (blue) are both conserved in time.\footnote{For internal energy we used the Debye model for copper, see e.g. \cite{landau5}.}
}
	\end{subfigure}
	\caption{\label{fig.1dt}
	Evolution of angular momentum, energy and magnitude of angular momentum when taking $\tau = \diff t$. Numerical results were obtained with scheme \eqref{eq.mx.tau} for $\tau = \diff t$. The scheme is non-dissipative as anticipated.}
\end{figure}

Let us set moments of inertia\footnote{the same as in \cite{Materassi} just with $x$ and $z$ swapped} to $I_x=1$, $I_y=5$, $I_z = 10$, time step $\diff t = 0.01$  and initial condition $m_x=0.1$, $m_y = 1.0$ and $m_z = 0.1$, i.e. in the unstable rotation around the middle axis. The relaxation time is varied between $\tau = 0$, $\tau = \diff t$ and $\tau = 100\diff t$. For $\tau = \diff t$ the scheme is non-dissipative as anticipated, see Fig. \ref{fig.m.1dt}.

\begin{figure}[ht]
	\centering
	\begin{subfigure}{.45\textwidth}
		\centering
		\includegraphics[width=0.95\linewidth]{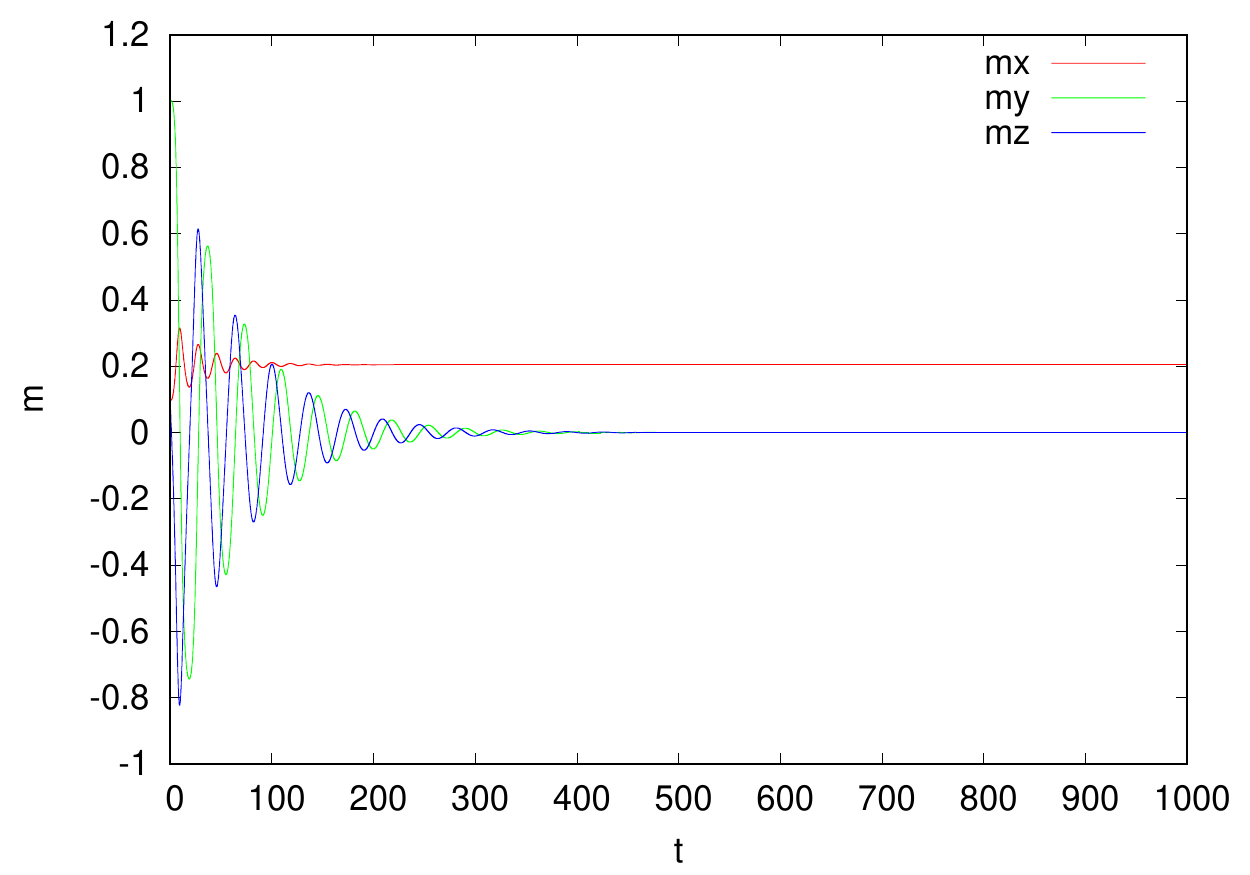}
		\caption{\label{fig.m.100dt}
		Components ($m_x$ red, $m_y$ green and $m_z$ blue) of angular momentum in time. The rotation starts as being aligned mostly in direction of the $y-$axis, which corresponds to the intermediate moment of inertia and is thus unstable. Finally it ends up in pure rotation around the $x-$axis, since it has lower $\mm^2$.}
	\end{subfigure}%
	\hspace{5pt}\begin{subfigure}{.45\textwidth}
		\centering
		\includegraphics[width=0.95\linewidth]{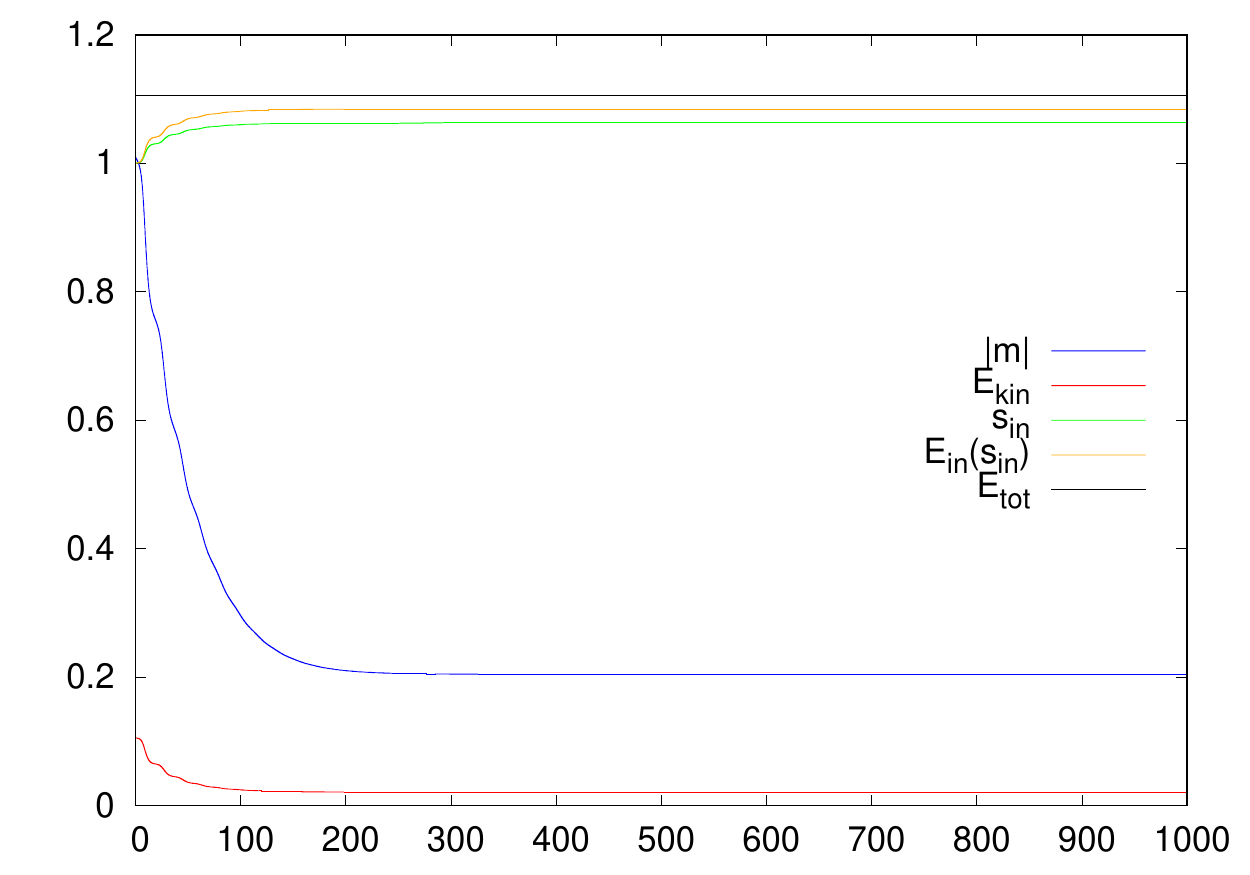}
		\caption{\label{fig.Em.100dt}
		Kinetic energy (red) and magnitude of angular momentum (blue)  are both reduced in time.}
	\end{subfigure}
	\caption{\label{fig.100dt}
	Evolution of angular momentum, energy and magnitude of angular momentum when taking $\tau = 100 \diff t$ obtained with scheme \eqref{eq.mx.tau}. The scheme is dissipative as expected. Total energy is conserved while kinetic energy is reduced. Internal entropy grows.}
\end{figure}

Taking $\tau = 100\diff t$, the scheme becomes dissipative as expected, see Fig. \ref{fig.100dt}, and energy is reduced as well as the magnitude of angular momentum.

\begin{figure}[ht]
	\centering
	\begin{subfigure}{.45\textwidth}
		\centering
		\includegraphics[width=0.95\linewidth]{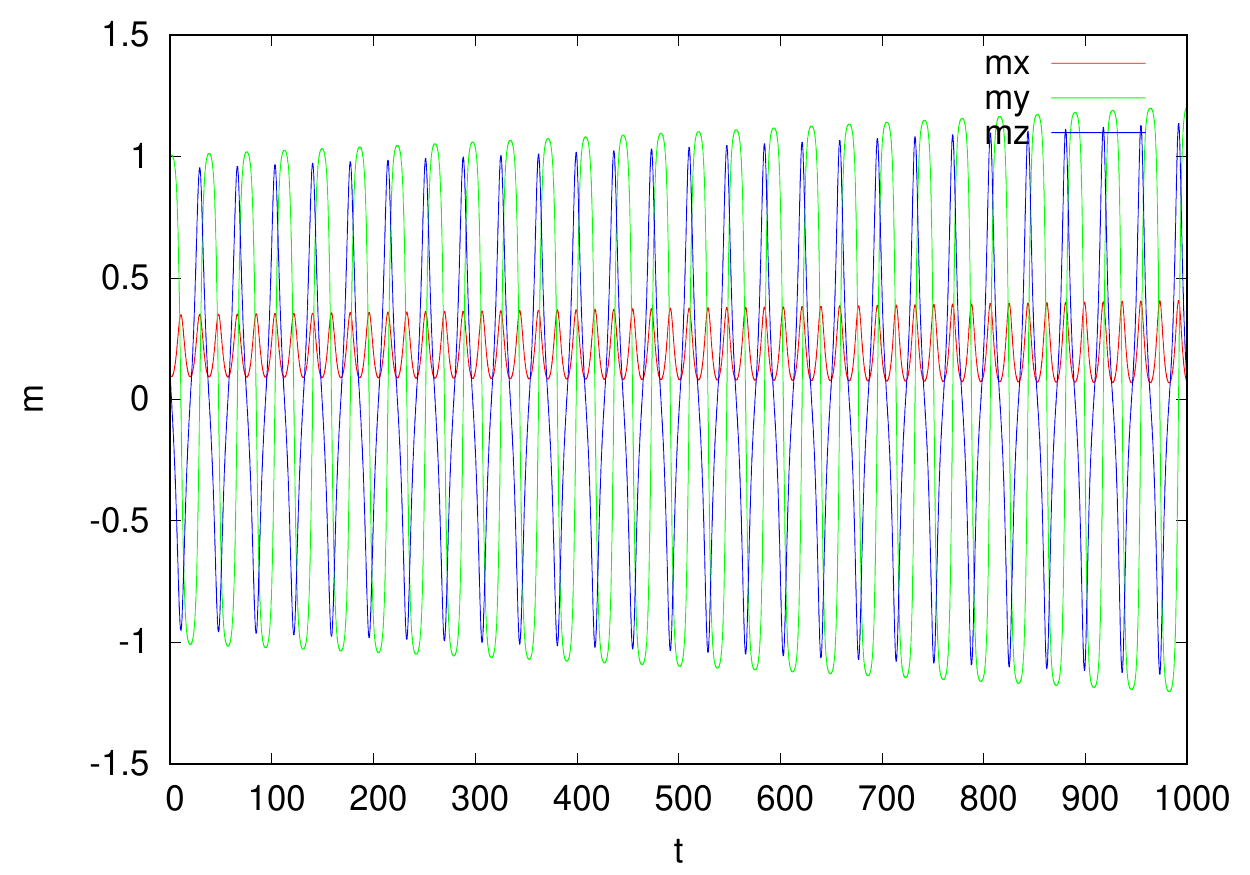}
		\caption{\label{fig.m.0dt}
		Components ($m_x$ red, $m_y$ green and $m_z$ blue) of angular momentum in time. The rotation starts as being aligned mostly in direction of the $y-$axis, which corresponds to the intermediate moment of inertia. }
	\end{subfigure}%
	\hspace{5pt}\begin{subfigure}{.45\textwidth}
		\centering
		\includegraphics[width=0.95\linewidth]{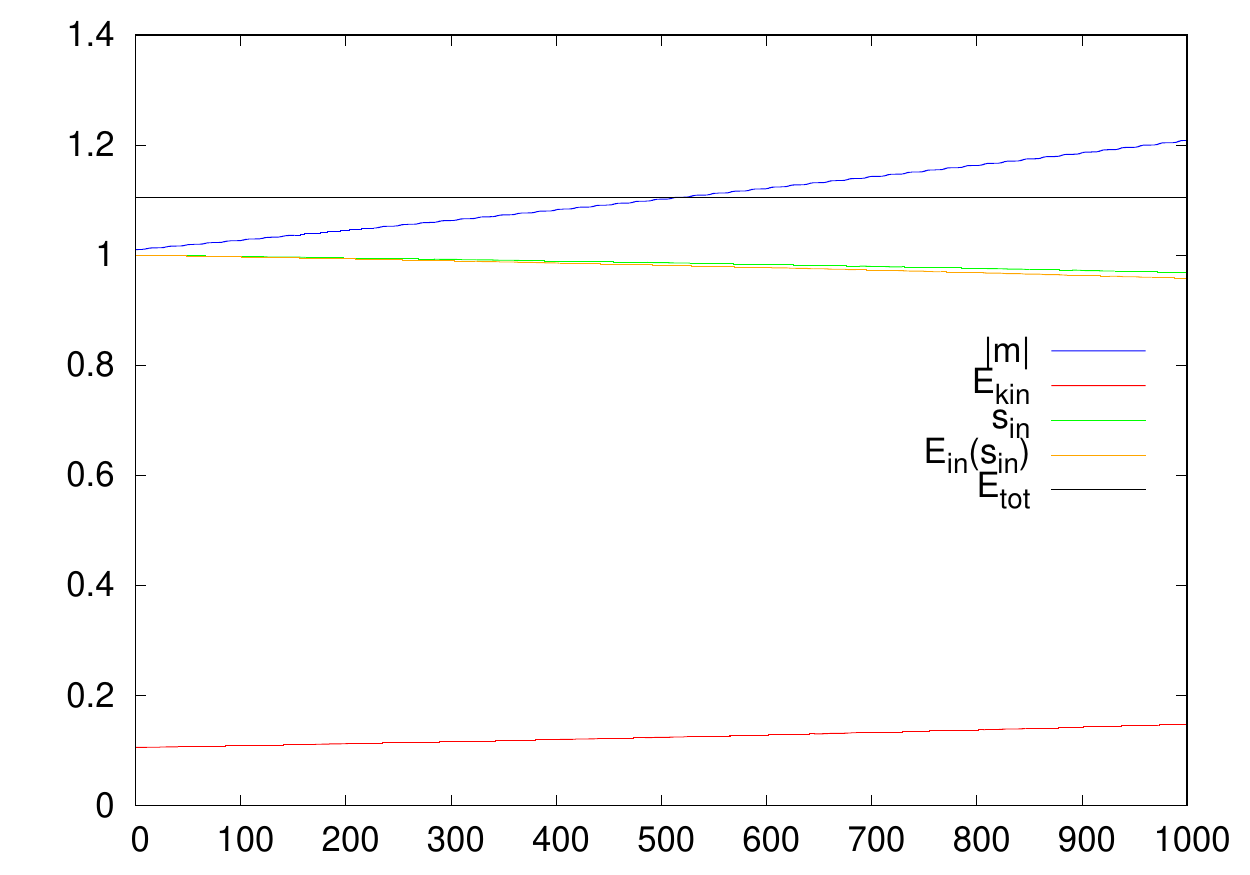}
		\caption{\label{fig.Em.0dt}
		Kinetic energy (red) and magnitude of angular momentum (blue) both grow in time.}
	\end{subfigure}
	\caption{\label{fig.0dt}
	Evolution of angular momentum, energy and magnitude of angular momentum when taking $\tau = 0$. Numerical results were obtained with scheme \eqref{eq.mx.tau}, which in this case turns to first-order forward Euler scheme for the reversible equations. The scheme is anti-dissipative as anticipated, which is nonphysical.}
\end{figure}

Taking $\tau = 0$, the scheme becomes anti-dissipative as expected, see Fig. \ref{fig.0dt}, and energy is raised as well as the magnitude of angular momentum.

\subsubsection{Stability by Energy-Casimir method}
It was shown in Sec. \ref{sec.SO3.stab} that the reversible evolution of rigid body rotation is stable when rotating around the minor or major axes ($x$ or $z$). The potentials were concave in the former case while convex in the latter case.

As discussed in Sec. \ref{sec.ECM.dis},  both rotations around the $z$-axis and $x-$axis become even nearly asymptotic stable when using the EhRe equations.

\subsubsection{Instability by eigenvalues}
Let us now discuss stability of linearized regularized evolution equations. Taking $\mm^0 = (m^0_x, 0, 0)$ and $\mm = \mm^0 + \delta \mm$, Eqs.  \eqref{eq.m.sr} become (after dropping all terms of the order $\Obig(\delta \mm)^2$)
\begin{equation}
	\frac{\diff}{\diff t}
	\begin{pmatrix}
		\delta m_x\\
		\delta m_y\\
		\delta m_z
	\end{pmatrix}
	=
	m^0_x \cdot
	\begin{pmatrix}
		0 & 0 & 0\\
		0 & \frac{\tau}{2} m^0_x J_y J_z & Jy\\
		0 & J_x & \frac{\tau}{2} m^0_x J_y J_z
	\end{pmatrix}.
\end{equation}
The nontrivial eigenvalues of the matrix on the right hand side are given by
\begin{equation}
	\lambda_{\pm} = \frac{\tau}{2} m^0_x J_y J_z \pm i \sqrt{-J_y J_z},
\end{equation}
since $J_y J_z<0$, and they thus have only negative real part.

Similarly, taking $\mm^0 = (0, 0, m^0_z)$ leads only to eigenvalues with only negative real parts when neglecting the zero eigenvalue.

Finally, $\mm^0 = (0, m^0_y, 0)$ leads to
\begin{equation}
	\lambda_{\pm} = \frac{\tau}{2} m^0_y J_z J_x \pm \sqrt{J_x J_z}.
\end{equation}
Since $J_x J_z>0$, the $\lambda_+$ eigenvalue is always positive entailing instability. 

In summary, analysis of eigenvalues of the linearized regularized equations tells that the rotation around the middle axis ($y$) is unstable.

\subsection{Energetic Ehrenfest regularization}
The EhRe evolution equations for freely rotating rigid body caused kinetic energy to dissipate and magnitude of angular momentum to decay. The latter property, however, is not desirable, since angular momentum should be conserved regardless the dissipation. Therefore, it is not the full EhRe evolution that should be taken as Ehrenfest regularization of the Hamiltonian rigid body motion, but the E-EhRe evolution, which conserves Casimirs (thus also magnitude of angular momentum) of the Poisson bracket while still dissipating kinetic energy.

\subsubsection{Energetic Ehrenfest-regularized evolution equations}
The energetic Ehrenfest regularization \eqref{eq.EhRe.M} becomes  in the case of Poisson bracket \eqref{eq.PB}
\begin{equation}\label{eq.SO3.M}
	\dot{m}_x = m_y m_z J_x -\frac{\tau}{2} m_x \left(m^2_z \frac{J_y}{I_y} - m^2_y \frac{J_z}{I_z}\right),
\end{equation}
where analogical evolution equations for $m_y$ and $m_z$ are obtained by cyclic permutation.

These equations reduce the energy while keeping the entropy constant. This in fact means that the magnitude of angular momentum $\mm^2$ is kept constant while reducing kinetic energy of the rotation. This is actually the sought physical behavior, speaking in favor of the energetic regularization instead of the full (EhRe) or entropic regularization (S-EhRe).

\subsubsection{Stability analysis}
The Energy-Casimir method for the E-EhRe evolution implies that pure rotation around the $z-$axis is stable and nearly asymptotically stable while not telling anything about pure rotation around the $x-$axis. Moreover, pure rotation around the $z-$ axis is the state with lowest kinetic energy for given value of $\mm^2$. By choosing the value of $\mm^2$, the only possible state of pure rotation around the $z-$axis, $\xx_0 = (0, 0, \sqrt{\mm^2})$, is already determined. That point is actually even asymptotically stable, $\mm\stackrel{t\rightarrow\infty}{\rightarrow}(0, 0, \sqrt{\mm^2})$, and the only stable state.

\subsubsection{Numerical solution}
\begin{figure}
	\centering
	\includegraphics{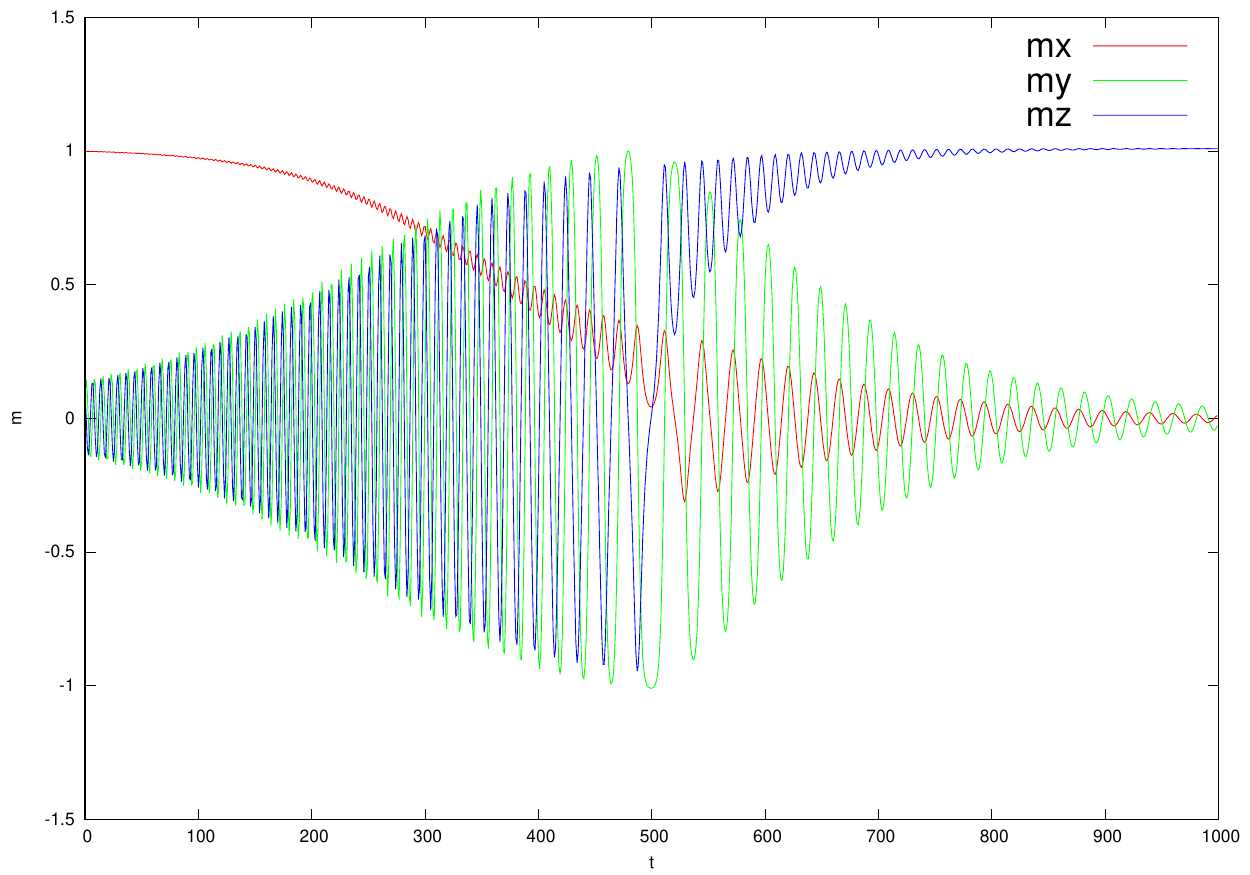}
	\caption{\label{fig.EEhRe.mx}Energetic Ehrenfest regularization (E-EhRe) of rigid body motion. Only pure rotation around the $z-$axis is stable (even asymptotically stable). Pure rotation around the minor $x-$axis becomes unstable as it has higher kinetic energy for given $\mm^2$ than pure rotation around the major $z-$axis.}
\end{figure}

The Crank-Nicolson discretization of Eq. \eqref{eq.SO3.M}, Eq. \eqref{eq.EhRe.M.CN}, leads to a numerical scheme that conserves Casimirs (up to the second order) and reduced kinetic energy. Results for the same setting as in Sec. \ref{sec.SO3.EhRe.num} and initial condition $\mm=(1.0, 0.1, 0.1)$, i.e. rotation nearly around the minor axis, are shown in Fig. \ref{fig.EEhRe.mx}.

Rotation around the middle axis is unstable (as for the reversible evolution or EhRe). Rotation around the minor axis becomes unstable as well and kinetic energy is dissipated while keeping magnitude of angular momentum constant. The final state is the rotation around the major axis. This is the sought physical behavior.

\subsection{Entropic Ehrenfest regularization}
For the sake of completeness, let us discuss also the entropic Ehrenfest regularization of the rigid body motion (S-EhRe). This evolution, however, exhibits nonphysical behavior, since it reduces the magnitude of angular momentum while preserving kinetic energy.

\subsubsection{Evolution equations}
The general form of the S-EhRe equations is given in Eq. \eqref{eq.EhRe.N}. The irreversible $\NN-$operator is calculated in accordance to Eq. \eqref{eq.Nij},
\begin{eqnarray}
	N^{ij} &=& -m_c\eps_{cab} \frac{\partial (-m_k \eps_{kij})}{\partial m_a}E_{m_b} \nonumber\\
	&=& m_c\eps_{ckb} \eps_{kij}E_{m_b}  = -m_c E_{m_b}(\delta_{ci}\delta_{bj}-\delta_{cj}\delta_{bi})\nonumber\\
	&=&-m_i E_{m_j}+m_j E_{m_i}.
\end{eqnarray}
In particular, $N^{xx} = 0$, $N^{xy} = m_x m_y (1/I_x - 1/I_y)=-m_x m_y J_z$ and $N^{xz}=-N^{zx} = m_z m_x J_y$. Other elements of the matrix are obtained by cyclic permutation. The final evolution equation \eqref{eq.EhRe.N} with state variables, Poisson bracket and energy for the rigid body rotations then reads
\begin{equation}\label{eq.SO3.ReEhRe}
	\dot{m}_x = m_y m_z J_x +\frac{\tau}{2} m_x \left(\frac{m^2_y}{I_y}\left(\frac{1}{I_x}-\frac{1}{I_y}\right)+\frac{m^2_z}{I_z}\left(\frac{1}{I_x}-\frac{1}{I_z}\right)\right).
\end{equation}
Note that the irreversible term on the right hand side and the irreversible term on the right hand side of Eq. \eqref{eq.SO3.M} sum up to the irreversible term in the EhRe evolution equation \eqref{eq.x.evo.sr}.
Equations \eqref{eq.SO3.ReEhRe} are the entropic regularized evolution equations for the rigid body rotation, which conserve energy while raising kinetic entropy (concave Casimirs).

\subsubsection{Comparison with literature}\label{sec.compare}
Evolution equation for rigid body rotation enhanced by adding dissipative terms with particularly advantageous properties were suggested in \cite{Materassi} recently. The dissipative terms represented a torque force on the rigid body and produced entropy while keeping energy constant. These properties are in fact the same as the entropic regularized evolution possesses, and we will show that the equations from \cite{Materassi} are in fact equivalent to the S-EhRe equations for rigid body rotation.

Denoting $E_\mm=\oomega$ as the angular velocity, the S-EhRe evolution equation \eqref{eq.SO3.ReEhRe} can be rewritten as
\begin{eqnarray}\label{eq.m.omega}
	\dot{\mm} &=& \mm\times \oomega +\frac{\tau}{2}\NN\cdot\omega\\
	&=&\mm\times\oomega - \frac{\tau}{2} \left(\oomega^2\Id -\oomega\otimes\oomega\right)\cdot\mm.
\end{eqnarray}
Choosing entropy $S(\mm) = \eta(\mm^2)$, $\eta$ being a concave function, as in \cite{Materassi}, the evolution equation becomes
\begin{equation}\label{eq.m.Maserati}
	\dot{\mm}
	=\mm\times\oomega - \frac{\tau}{\eta'(\mm)} \underbrace{\left(\oomega^2\Id -\oomega\otimes\oomega\right)}_{=\GGamma}\cdot S_\mm,
\end{equation}
where $\GGamma$ is a symmetric positive semidefinite matrix (as follows from the Cauchy-Schwarz inequality). Concavity of entropy implies in particular that $\eta'\leq 0$, as shown in Sec. \ref{sec.SO3.stab}, and the coefficient in front of the matrix is thus always positive. The equation can be thus seen as a metriplectic system or as GENERIC. However, the behavior predicted by
the S-EhRe evolution for rigid body is nonphysical, since it implies asymptotic stability of rotation around the minor (not major) axis.

Note that it is actually natural that entropy emerges from the Poisson bivector as when rewriting Eq. \eqref{eq.m.omega} to Eq. \eqref{eq.m.Maserati}, since entropy is a Casimir of the Poisson bracket. The Lie-Poisson evolution takes place on symplectic leaves, each of which is characterized by some values of the Casimirs, e.g. entropy. Entropy is already encoded into the mechanics in this sense.

\subsubsection{Stability by Energy-Casimir method}
A feature of the evolution equations is that entropy (a concave Casimir) grows while energy is kept constant. Let us first analyze stationary points of the dynamics from the perspective of energy and entropy. Considering a point $\xx_0 = (0, 0, m_z^0)$, i.e. rotation around the major axis, energy and entropy of the rotation are $E^0=\frac{(m^0_z)^2}{2I_z}$ and $S=-\eta((m_z^0)^2)$. Consider now another point $(m_x, m_y, m_z)$ where the system was before reaching $\xx^0$.  The point has energy \eqref{eq.SO3.E}, which is equal to the energy of pure rotation around the $z-axis$,
\begin{equation}
	\frac{1}{2}\left(\frac{(m_x)^2}{2I_x}+\frac{(m_y)^2}{2I_y}+\frac{(m_z)^2}{2I_z}\right) = \frac{(m^0_z)^2}{2I_z}.
\end{equation}
Since the system was in that point before approaching $\xx_0$, entropy of the system must be lower at that point than at $\xx_0$, which means
\begin{equation}
	(m_x^0)^2 \leq (m_x)^2 + (m_y)^2 + (m_z)^2,
\end{equation}
due to the concavity of entropy. This inequality means that
\begin{equation}
 \frac{(m^0_z)^2}{2I_z}=
	 \frac{1}{2}\left(\frac{(m_x)^2}{2I_x}\frac{I_x}{I_z}+\frac{(m_y)^2}{2I_y}\frac{I_y}{I_z}+\frac{(m_z)^2}{2I_z}\right) >
\frac{(m^0_z)^2}{2I_z},
\end{equation}
since $I_x < I_y < I_z$, which is a contradiction. Therefore, the state of pure rotation around the $z-$axis can not be reached by the entropic EhRe, since it has lower (kinetic) entropy than other points.

This is also why in \cite{Materassi} relaxation towards to $z-$axis, which is of course the most physically reasonable, is obtained only either when taking negative phenomenological coefficient (corresponding to our $\tau$) or be taking convex entropy.

Also the Energy-Casimir method only indicates stability of rotation around the $x-$axis while not telling anything (or indicating instability) of pure rotation  around the $z-$axis. The E-EhRe evolution should be preferred to the S-EhRe
or EhRe evolutions for the rigid body rotations.

\section{Fluid mechanics}
Let the set of state variables be density, momentum density and entropy density, $\xx = (\rho, \uu, s)$. This is the level of description of fluid mechanics. Mechanics of fluids is Hamiltonian and is expressed by the Poisson bracket (see e.g. \cite{marsden-weinstein})
\begin{multline}
	\{F,G\} = \int \diff\rr \rho (\partial_i F_\rho G_{u_i} - \partial_i G_\rho F_{u_i})\\
	+\int \diff\rr u_i (\partial_j F_{u_i} G_{u_j} - \partial_j G_{u_i} F_{u_j})\\
	+\int \diff\rr s (\partial_i F_s G_{u_i} - \partial_i G_s F_{u_i}).
\end{multline}
Substituting energy for the functional $G$ and rewriting this bracket into the form
\begin{equation}
	\{F,E\} = \int\diff \rr A_\rho\cdot(\dots)
	+ \int\diff \rr A_{u_i}\cdot(\dots)
	+ \int\diff \rr A_s\cdot(\dots),
\end{equation}
the evolution equations for state variables $(\rho,\uu,s)$ can be easily read.
The evolution equations, the Euler equations for compressible fluids, are
\begin{subequations}\label{eq.Euler}
\begin{eqnarray}
	\partial_t \rho &=& -\partial_i (\rho E_{u_i})\\
	\partial_t u_i &=& -\rho \partial_i E_\rho -u_j \partial_i E_{u_j} - s \partial_i E_s - \partial_j(u_i E_{u_j})\\
	 &=& -\partial_i p - \partial_j(u_i E_{u_j})\nonumber\\
	\partial_t s &=& -\partial_i (s E_{u_i}).
\end{eqnarray}
\end{subequations}
Note that pressure can be in general identified as
\begin{equation}
	p = -e + \rho E_\rho + u_i E_{u_i} + s E_s,
\end{equation}
where $e$ is the total energy density, typically
\begin{equation}\label{eq.E.Euler}
	E = \int \diff \rr \underbrace{\frac{\uu^2}{2\rho} + \int \diff \rr \eps(\rho,s)}_{=e}.
\end{equation}
The first part is the kinetic energy and the second is the internal energy. The expression for pressure can also be simplified to
\begin{equation}\label{eq.p}
	p = -\eps + \rho \eps_\rho + s\eps_s.
\end{equation}
which is compatible with the usual notion of pressure in local thermodynamic equilibrium \cite{dgm}. Evolution equations \eqref{eq.Euler} express the Hamiltonian evolution of fluids.

The Euler evolution equations are reversible with respect to time-reversal transformation and they obviously do not produce entropy, they are non-dissipative. Let us now demonstrate the regularization of the equations.

\subsection{Ehrenfest regularization}
The dissipative terms in the general equation for Hamiltonian Ehrenfest regularization, Eq. \eqref{eq.x.evo.sr}, can be expressed as
\begin{equation}
	\frac{\tau}{2} \{\{x^i, E\}, E\}
	=\frac{\tau}{2} \{\dot{x}^i_{\rev}, E\},
\end{equation}
where $\dot{x}^i_{\rev}$ is the Hamiltonian part of the evolution.

In particular, regularization of density evolution
\begin{subequations}
\begin{equation}
	\{(\partial_t \rho)_\rev,E\}  = \left\{ - \int \dr' \delta(\rr-\rr') \partial'_i (\rho(\rr') E_{u_i}(\rr')), E\right\},
\end{equation}
where $\partial'_i$ stands for $\partial/\partial r'^i$,
requires calculation
\begin{equation*}
	\frac{\delta}{\delta \rho(\rr')} \int \dr' \delta(\rr-\rr') \partial'_i (\rho E_{u_i})|_{\rr'}
	= -\frac{\delta}{\delta \rho(\rr)} \int \dr' \partial'_i \delta(\rr-\rr') u_i(\rr') = 0.
\end{equation*}
Similarly
\begin{align}
	-\frac{\delta}{\delta u_i(\rr')} \int \dr' \delta(\rr-\rr') \partial'_j u_j(\rr') &= \partial'_i \delta(\rr-\rr'),\\
	\frac{\delta}{\delta s(\rr')} \int \dr' \delta(\rr-\rr') \partial'_i u_i(\rr') &= 0,
\end{align}
\end{subequations}
and hence the regularization entails an irreversible correction to the evolution equation in the form
\begin{eqnarray}
	\{(\partial_t \rho(\rr))_\rev,E\} &=& \int \dr' \rho(\rr') \left(-\partial'_k E_\rho \partial'_k \delta(\rr-\rr')\right)\nonumber\\
	&&+\int\dr' u_k(\rr')\left(\partial'_l \partial'_k \delta(\rr-\rr') E_{u_l}(\rr') - \partial'_l E_{u_k}(\rr') \partial'_l \delta(\rr-\rr')\right)\nonumber\\
	&&+\int\dr' s(\rr')\left(-\partial'_k E_s \partial'_k \delta(\rr-\rr')\right)\nonumber\\
	&=& \partial_k \partial_k p + \partial_k\partial_l \left(\frac{u_k u_l}{\rho}\right).
\end{eqnarray}
The regularization thus adds a dissipative flux to the equation for density (to the mass balance equation).

Let us now determine entropy regularization. Variations
\begin{subequations}
\begin{eqnarray}
	-\frac{\delta}{\delta \rho(\rr')} \int \dr' \delta(\rr-\rr') \partial'_i (s u_i/\rho)|_{\rr'} &=&
	- \frac{s u_i}{\rho^2}\Big|_{\rr'} \partial'_i \delta(\rr-\rr'),\\
	-\frac{\delta}{\delta u_i(\rr')} \int \dr' \delta(\rr-\rr') \partial'_i (s u_i/\rho)|_{\rr'} &=&
	\partial_i \delta(\rr-\rr') \frac{s(\rr')}{\rho(\rr')},\\
	-\frac{\delta}{\delta s(\rr')} \int \dr' \delta(\rr-\rr') \partial'_i (s u_i/\rho)|_{\rr'} &=&
	\partial'_i \delta(\rr-\rr') \frac{u_i}{\rho}\Big|_{(\rr')},
\end{eqnarray}
\end{subequations}
result in regularization
\begin{equation}
	\{\partial_t s, E\} = \partial_k \left(\frac{s}{\rho} \partial_k p\right) + \partial_k\partial_l\left( s\frac{u_k}{\rho}\frac{u_l}{\rho}\right).
\end{equation}
A dissipative entropy flux thus appears.

Finally regularization of momentum follows from variations
\begin{subequations}
\begin{eqnarray}
	\frac{\delta}{\delta \rho(\rr')} \int \dr' \delta(\rr-\rr') \left(-\partial'_i p -\partial'_j(u_i u_j/\rho)\right)\Big|_{\rr'} &=&
	\partial'_i \delta(\rr-\rr') p_\rho(\rr') \nonumber\\
	&&- \partial'_j \delta(\rr-\rr') \frac{u_i}{\rho}\frac{u_j}{\rho}\Big|_{\rr'}\\
	\frac{\delta}{\delta u_k(\rr')} \int \dr' \delta(\rr-\rr') \left(-\partial'_i p -\partial'_j(u_i u_j/\rho)\right)\Big|_{\rr'} &=&
	\delta_{ik} \frac{u_j}{\rho} \partial'_j \delta(\rr-\rr') + \frac{u_i}{\rho} \partial'_k \delta(\rr-\rr')\\
	\frac{\delta}{\delta s(\rr')} \int \dr' \delta(\rr-\rr') \left(-\partial'_i p -\partial'_j(u_i u_j/\rho)\right)\Big|_{\rr'} &=&
	p_s(\rr') \partial'_i \delta(\rr-\rr'),
\end{eqnarray}
where $p_\rho$ and $p_s$ stand for derivatives of pressure \eqref{eq.p} with respect to density and entropy, respectively.
\end{subequations}
After some algebra we obtain
\begin{eqnarray}
	\{(\partial_t u_i)_\rev, E\} &=& \partial_j \left(\frac{u_j}{\rho} \partial_i p + \frac{u_i}{\rho} \partial_j p\right)
	+ \partial_i \left( \partial_k u_k p_\rho + \partial_k\left(\frac{u_k s}{\rho}\right) p_s\right)\nonumber\\
	&&+\partial^2_{j,k} \left(\frac{u_i u_j u_k}{\rho^2}\right)
	.
\end{eqnarray}
The regularization thus also brings new dissipative fluxes of momentum.

In summary, the regularized Euler evolution equations become
\begin{subequations}\label{eq.FM.evo.sr}
	\begin{eqnarray}
		\label{eq.rho.sr}		\partial_t \rho &=& -\partial_j (\rho v_j) -\partial_j \left(-\frac{\tau}{2}\partial_j p - \frac{\tau}{2}\partial_k (\rho v_j v_k)\right)\\
		\label{eq.ui.sr}\partial_t u_i &=& -\partial_i p -\partial_j(\rho v_i v_j)\\
		&&-\partial_j \left(-\frac{\tau}{2}\left(v_j \partial_i p + v_i \partial_j p\right)
                   - \frac{\tau}{2}\partial_k (\rho v_i v_j v_k)
		   \right)- \partial_i \left(-\frac{\tau}{2}\left( \partial_k (\rho v_k) p_\rho + \partial_k(s v_k) p_s\right)\right)\nonumber\\
		\label{eq.s.sr}\partial_t s &=& -\partial_j (s v_j)
		-\partial_j\left(-\frac{\tau}{2} \frac{s}{\rho}\partial_j p - \frac{\tau}{2}\partial_k (s v_j v_k)\right),
	\end{eqnarray}
	where velocity was identified as $v_i = u_i /\rho$.
\end{subequations}
The first part of the irreversible flux on the right hand side of Eq. \eqref{eq.rho.sr} resembles Darcy's law while the second part is a higher-order contribution from spatial variations of velocity. Mass is, of course, conserved. The irreversible flux in Eq. \eqref{eq.ui.sr}, the dissipative stress tensor, is symmetric. Momentum as well as angular momentum are thus conserved. The irreversible entropy flux also contains a Darcy-like contribution as well as contribution from velocity variations. Entropy is also conserved. We conjecture that numerical solution of these equations with forward Euler scheme $\tau=\diff t$ and a finite element spatial discretization should lead to viscous-like solutions to the compressible Euler equations.

\subsection{Entropic Ehrenfest regularization}
Instead of the full EhRe regularization, one can carry out the entropic regularization (S-EhRe) of the Euler equations \eqref{eq.Euler}. The $\NN-$operator then reads
\begin{subequations}
	\begin{eqnarray}
	N^{\rho(\rr_a),u^i(\rr_b)} &=& -\frac{\partial}{\partial r_b^i} \delta(\rr_b-\rr_a) \frac{\partial}{\partial r_b^j} (\rho(\rr_b)E_{u^j}(\rr_b)),\\
	N^{u^i(\rr_a),s(\rr_b)} &=& \frac{\partial}{\partial r_a^i} \delta(\rr_b-\rr_a) \frac{\partial}{\partial r_a^j} (s(\rr_a)E_{u^j}(\rr_a)),\\
	N^{u^i(\rr_a),u^j(\rr_b)} &=& -\rho(\rr_b) \frac{\partial}{\partial r_b^i} E_\rho(\rr_b)  \frac{\partial}{\partial r_b^j} \delta(\rr_b-\rr_a) +\rho(\rr_a) \frac{\partial}{\partial r_a^j} E_\rho(\rr_a)  \frac{\partial}{\partial r_a^i} \delta(\rr_b-\rr_a)\\
	&&-\frac{\partial}{\partial r_b^l} (u^i E_{u^l}(\rr_b))  \frac{\partial}{\partial r_b^j} \delta(\rr_b-\rr_a)+\frac{\partial}{\partial r_a^l} (u^j E_{u^l}(\rr_a))  \frac{\partial}{\partial r_a^i} \delta(\rr_b-\rr_a)\nonumber\\
		&& -u^k(\rr_b)\frac{\partial}{\partial r_b^i} (E_{u^k}(\rr_b))  \frac{\partial}{\partial r_b^j} \delta(\rr_b-\rr_a) + u^k(\rr_a)\frac{\partial}{\partial r_a^j} (E_{u^k}(\rr_a))  \frac{\partial}{\partial r_a^i} \delta(\rr_b-\rr_a)\nonumber\\
	&&-s(\rr_b)\frac{\partial}{\partial r_b^i} (E_{s}(\rr_b))  \frac{\partial}{\partial r_b^j} \delta(\rr_b-\rr_a)+s(\rr_a)\frac{\partial}{\partial r_a^j} (E_{s}(\rr_a))  \frac{\partial}{\partial r_a^i} \delta(\rr_b-\rr_a).\nonumber
	\end{eqnarray}
\end{subequations}
The regularized evolution equations are then
\begin{subequations}
\begin{eqnarray}
  \partial_t \rho &=& \ldots + \frac{\tau}{2} \partial_i\left(\partial_j(\rho E_{u^j}) E_{u^i}\right)\\
  \partial_t u_i &=& \ldots +\frac{\tau}{2} \Bigg[\partial_j u^j \partial_i E_\rho+\partial_j\left(\frac{s u^j}{\rho}\right)\partial_i E_s +\rho \partial_j E_\rho \partial_i\left(\frac{u^j}{\rho}\right)+\partial_j \left( \partial_i E_\rho u^j\right)\nonumber\\
                  &+& \partial_k\left(\frac{u^j u^k}{\rho}\right)\partial_i\left(\frac{u^j}{\rho}\right)+\partial_j \left(\partial_k\left(\frac{u^i u^k}{\rho}\right)\frac{u^j}{\rho}\right)+u^k\partial_j\left(\frac{u^k}{\rho}\right)\partial_i\left(\frac{u^j}{\rho}\right)\nonumber\\
  &+& \partial_j \left(\frac{u^k u^j}{\rho}\partial_i\left(\frac{u^k}{\rho}\right)\right) + s \partial_j E_s \partial_i\left(\frac{u^j}{\rho}\right)+\partial_j\left(s \partial_i E_s \frac{u^j}{\rho}\right)\Bigg],\\
 \partial_t s &=& \ldots + \frac{\tau}{2} \partial_i\left(\partial_j(s E_{u^j}) E_{u^i}\right),
\end{eqnarray}
\end{subequations}
where $\ldots$ stands for the corresponding reversible part of the evolution, i.e. right hand side of Eqs. \eqref{eq.Euler}.

Using energy \eqref{eq.E.Euler}, these evolution equations can be rewritten to
\begin{subequations}
	\begin{eqnarray}
	\partial_t \rho &=& \ldots + \frac{\tau}{2} \partial_i \left(v_i \partial_j (\rho v_j)\right)\\
	\partial_t u_i &=& \ldots + \frac{\tau}{2}\Big[ \partial_j(\partial_k(\rho v_i v_k)v_j) + \rho v_k \partial_k v_j \partial_i v_j\nonumber\\
	&+&\partial_j(v_j\partial_i p) + \partial_j v_j \partial_i p + \partial_i v_j \partial_j p\nonumber\\
	&+& v_j\left(\partial_j \rho \partial_i \eps_\rho + \partial_j s\partial_i \eps_s\right)\Big]\label{eq.uSEhRe}\\
	\partial_t s &=& \ldots + \frac{\tau}{2} \partial_i \left(v_i \partial_j (\rho s_j)\right)
	\end{eqnarray}
\end{subequations}
Mass and entropy are conserved in the S-EhRe evolution while changing total momentum.

Note that without accounting for the influence of entropy we may rewrite equation for momentum \eqref{eq.uSEhRe}
\begin{equation} \label{eq.SEhRe.ui-woS}
  	\partial_t u_i = \ldots + \frac{\tau}{2}\Big[ \partial_j(\partial_k(\rho v_i v_k)v_j) + \rho v_k \partial_k v_j \partial_i v_j + \frac{1}{\rho}\partial_j\left(u_j\partial_i p \right)+ \partial_j v_j \partial_i p + \partial_i v_j \partial_j p\Big],
\end{equation}
        as $\partial_i p=\rho \partial_i \epsilon_\rho$.

\subsection{Energetic Ehrenfest regularization}

Energetic regularization does not conserve energy and hence in the case of Euler equations we add internal entropy to have a system conserving energy while total entropy is being increased. For this reason, we shall focus only on energetic regularization of density and momentum while neglecting changes due to (kinetic) entropy.

The symmetric operator $\MM$ has no trivial components and their explicit form is
\begin{equation*}
  M^{\rho(\rr_a)\rho(\rr_b)} = -\frac{\partial}{\partial r_a^i} \left(\rho(\rr_a)\frac{\partial}{\partial r_a^i} \delta(\rr_a-\rr_b)\right),
\end{equation*}
\begin{equation*}
  M^{\rho(\rr_a) u^i(\rr_b)} = \rho(\rr_b) \frac{\partial}{\partial r_b^i} \left(\frac{u^j(\rr_b)}{\rho(\rr_b)} \frac{\partial}{\partial r_b^j}\delta(\rr_b-\rr_a)\right)-u_j(\rr_b)\frac{\partial^2}{\partial r_b^i\partial r_b^j}\delta(\rr_b-\rr_a) - \frac{\partial}{\partial r_a^j}\left(u^i(\rr_a)\frac{\partial}{\partial r_a^j}\delta(\rr_b-\rr_a)\right)
\end{equation*}
\begin{align*}
  M^{u^j(\rr_a)u^k(\rr_b)}& = \rho(\rr_a) \rho(\rr_b) \frac{\partial^2}{\partial r_a^i\partial r_b^j}\left(E_{\rho\rho} \delta(\rr_b-\rr_a)\right)-\rho(\rr_a)u^k(\rr_b) \frac{\partial^2}{\partial r_a^i\partial r_b^j}\left(\frac{u^k}{\rho^2} \delta(\rr_b-\rr_a)\right)\\
                          &-\rho(\rr_a) \frac{\partial^2}{\partial r_a^i\partial r_b^k}\left(\frac{u^k u^j}{\rho^2}\delta(\rr_b-\rr_a)\right)-u^k(\rr_a)\rho(\rr_b)\frac{\partial^2}{\partial r_a^i\partial r_b^j}\left(\frac{u^k}{\rho^2}\delta(\rr_b-\rr_a)\right)\\
                          &+u^k(\rr_a)u^k(\rr_b)\frac{\partial^2}{\partial r_a^i\partial r_b^j}\left(\frac{1}{\rho} \delta(\rr_b-\rr_a)\right) - \rho(\rr_b) \frac{\partial^2}{\partial r_a^k\partial r_b^j}\left(\frac{u^k u^i}{\rho^2} \delta(\rr_b-\rr_a)\right)\\
  &+u^k(\rr_a) \frac{\partial^2}{\partial r_a^i\partial r_b^k}\left(\frac{u^j}{\rho} \delta(\rr_b-\rr_a)\right) + u^k(\rr_b) \frac{\partial^2}{\partial r_a^k \partial r_b^j}\left(\frac{u^i}{\rho}\delta(\rr_b-\rr_a)\right)\\
	&+ \frac{\partial^2}{\partial r_a^k\partial r_b^k}\left(\frac{u^i u^j}{\rho} \delta(\rr_b-\rr_a)\right).
\end{align*}

After rather tedious calculations the energetic regularized correction to Euler evolution equations can be revealed
\begin{eqnarray}
  \partial_t \rho &=& \ldots + \frac{\tau}{2}\left[\partial_k(\rho\partial_k \epsilon_\rho) + \partial_k\left(u^j \partial_j \frac{u^k}{\rho}\right)\right]=\ldots + \frac{\tau}{2}\left[\partial^2_{kk} p + \partial_k\left(u^j \partial_j \frac{u^k}{\rho}\right)\right]=\\
  \partial_t u_i &=& \ldots +\frac{\tau}{2}\left[\mathcal{W}^{ij} \partial_j E_\rho+\frac{u^j}{\rho} \partial_i \rho \partial_j E_\rho+u^i\partial^2_{jj} E_\rho+\rho\partial_i(\epsilon_{\rho\rho} \partial_j u^j)+\mathcal{W}^{ik} u^j \mathcal{D}^{jk} + \frac{u^i}{\rho} \partial_k(u^j \mathcal{D}^{jk})\right]\nonumber\\
                  &=& \ldots + \frac{\tau}{2} \Bigg[v^i\partial_k(\rho v^j \partial_j v^k)+\rho v^j \mathcal{W}^{ik} \partial_j v^k + \rho v^i \partial_j\left(\frac{1}{\rho} \partial_j p\right) +\frac{1}{\rho}\left(\partial_j (\rho v^i) \partial_j p - \partial_j(\rho v^j)\partial_i p\right)\nonumber\\
		  &&-\partial_i v^j \partial_j p +\partial_i (p_\rho \partial_j(\rho v^j))\Bigg],\nonumber
\end{eqnarray}
where $\mathbf{\mathcal{D}}$ denotes the symmetric velocity gradient, i.e. $\mathcal{D}^{ij}=\partial_j \frac{u^i}{\rho}+\partial_i \frac{u^j}{\rho}$, and $\mathbf{\mathcal{W}}$ denotes the antisymmetric velocity gradient, $\mathcal{W}^{ij}=\partial_j \frac{u^i}{\rho}-\partial_i \frac{u^j}{\rho}$.

The reader can check that indeed the sum of E-EhRe and S-EhRe of Euler evolution equations gives EhRe relationships when neglecting the entropy (consistent with the above equations for E-EhRe description) entailing relations $p_\rho = \rho \epsilon_{\rho\rho}$ and $\partial_i p = \rho \partial_i \epsilon_\rho$.

\subsection{Ehrenfest regularization with internal entropy}

The various above flavors of Ehrenfest regularizations can be summarized as:
\begin{enumerate}
\item EhRe does not increase entropy (which is linear, not concave) but takes away energy. Evolution equations reveal that momentum is conserved.
\item Entropic EhRe conserves energy but also conserves entropy due to its linearity.
\item Energetic EhRe consumes energy while conserving entropy. However, by inspection of the evolution equations one can observe that momentum is not conserved. By adding internal entropy $s_{in}$ the correct the energetic balance is obtained.
\end{enumerate}

By revisiting the regularization of EhRe evolution, \eqref{eq.s.evo.sre}, one reveals a simple relation to E-EhRe regularization:
\begin{equation*}
  \dot{s}_{in} = \frac{1}{E_{s_{in}}^{tot}} \frac{\tau}{2} L^{il} E_{x^l} \frac{\delta^2 E}{\delta x^i \delta x^j} L^{jk} E_{x^k} = \frac{1}{E_{s_{in}}^{tot}} \frac{\tau}{2}  E_{x^l} M^{lk} E_{x^k}.
\end{equation*}
Therefore the evolution of the new state variable, internal entropy $s_{in}$, can be obtained by multiplying the E-EhRe correction of the evolution equations for the state variable $x^i$ by the corresponding $E_{x^i}$ and summing over all state variables.

In summary, the version of Ehrenfest regularization most suitable for ideal hydrodynamics is the full EhRe with internal entropy.

\section{Kinetic theory}
In kinetic theory the role of state variables is played by the one-particle distribution function $f(t,\rr,\pp)$, which expresses probability that a particle with momentum $\pp$ is present at position $\rr$ at time $t$. The evolution for the distribution function is usually described by Boltzmann equation, see e.g. \cite{marsden-bbgky,hierarchy,dgm}. The reversible part of the evolution is, however, Hamiltonian and the Poisson bracket is actually a Lie-Poisson bracket. Therefore, the Ehrenfest regularization applies to that reversible evolution with all its consequences. Which form of the regularization should be used? Since we wish to conserve kinetic energy and produce Boltzmann entropy, which is a concave Casimir of the Poisson bracket, we choose the entropic regularization (S-EhRe). Note that normalization of the distribution function is not affected by the regularization, since average of the distribution function is a linear Casimir (and thus does not grow).

\subsection{Hamiltonian dynamics}
But let us first recall the Hamiltonian formulation of the reversible evolution of the distribution function (see e.g. \cite{PKG} for more details).
The Boltzmann Poisson bracket of any two functionals $F(f)$ and $G(f)$ is
\begin{equation}\label{eq.KT.PB}
	\{F,G\} = \int \dr\int \dpp f \left(\frac{\partial F_f}{\partial \rr}\cdot \frac{\partial G_f}{\partial \pp} - \frac{\partial G_f}{\partial \rr}\cdot \frac{\partial F_f}{\partial \pp}\right).
\end{equation}
This bracket is generated by Poisson bivector
\begin{equation}
	L^{f(\rr,\pp),f(\rr',\pp')} = \frac{\partial \delta(\pp-\pp')}{\partial p'_k}\frac{\partial \delta(\rr-\rr')f(\rr',\pp')}{\partial r'_k}-
    \frac{\partial \delta(\pp-\pp')}{\partial p_k}\frac{\partial \delta(\rr-\rr')f(\rr,\pp)}{\partial r_k},
\end{equation}
as shown for instance in \cite{pre2014}.

Assuming ideal gas, the energy is prescribed as\footnote{Note that for this energy E-EhRe does not give any irreversible term, and EhRe is then equivalent to S-EhRe.}
\begin{equation}
	E(f) = \int \dr\int \dpp \frac{\pp^2}{2m}f(\rr,\pp),
\end{equation}
and Boltzmann entropy can be expressed as
\begin{equation}
	S(f) = \int \dr\int \dpp \eta(f),
\end{equation}
where $\eta(f)$ is typically $-k_B (\ln (h^3 f)-1)$, see e.g. \cite{PKG}.
Concavity of $S(f)$ is fulfilled when $\eta$ is also a concave function of $f$, $\eta''\leq 0$.

The reversible evolution equation for the distribution function is
\begin{equation}
	\partial_t f = \{f,E\} = -\frac{p_i}{m}\frac{\partial f}{\partial r_i}.
\end{equation}
Let us now develop the entropic regularization of this equation.

\subsection{Entropic Ehrenfest regularization}
The $\NN-$operator is given by formula \eqref{eq.Nij}, explicitly by
\begin{equation}
	N^{f(\rr,\pp),f(\rr',\pp')} = -\frac{p_l}{m}\frac{\partial \delta(\rr-\rr')}{\partial r'_k} \frac{\partial \delta(\pp-\pp')}{\partial p_k} \frac{\partial f(\rr',\pp)}{\partial r'_l}
	+\frac{p'_l}{m}\frac{\partial \delta(\rr'-\rr)}{\partial r_k} \frac{\partial \delta(\pp-\pp')}{\partial p'_k} \frac{\partial f(\rr,\pp')}{\partial r_l}.
\end{equation}
The final S-EhRe evolution equation for the distribution function is then
\begin{eqnarray}\label{eq.KT.SEhRe}
	\partial_t f &=& -\frac{p_i}{m}\frac{\partial f}{\partial r_i} + \frac{\tau}{2} \int \dr\int\dpp N^{f(\rr,\pp),f(\rr',\pp')} E_{f(\rr',\pp')}\nonumber\\
	&=&-\frac{p_i}{m}\frac{\partial f}{\partial r_i} + \frac{\tau}{2} \frac{p_k}{m}\frac{p_l}{m} \frac{\partial^2 f}{\partial r_k \partial r_l}.
\end{eqnarray}
This evolution equation consists of a reversible (Hamiltonian) term and an irreversible term.

Let us now discuss properties of Eq. \eqref{eq.KT.SEhRe}. Firstly, kinetic energy is expected to be preserved. Indeed,
\begin{equation}
	\dot{E} = 0 + \frac{\tau}{2}\int\dr\int\dpp \frac{\pp^2}{2m} \frac{p_k}{m}\frac{p_l}{m}\frac{\partial^2 f}{\partial r_k \partial r_l} = 0,
\end{equation}
where the last equality comes from that the boundary terms disappear (isolated system or a torus). Kinetic energy is thus conserved. Secondly, entropy (a concave Casimir of the Poisson bracket) grows,
\begin{equation}
	\dot{S} = 0 - \frac{\tau}{2} \int \dr\int\dpp \frac{p_k}{m} \frac{\partial f}{\partial r_k}\frac{p_l}{m} \frac{\partial f}{\partial r_l} \eta''(f)\geq 0,
\end{equation}
as expected. Boltzmann (kinetic) entropy is thus produced by the S-EhRe.

Finally, note Eq. \eqref{eq.KT.SEhRe} contains a diffusion-like term on the right hand side, which means that tendency to spatial homogenization can be anticipated. In other words, by ``smoothing out'' the solutions to the original Vlasov equation, a tendency to spatial homogenization may emerge. But this is actually how one can see Landau damping \cite{landau-damping,villaniv,KRM,PKG-Landau}, where irregularities of solutions build up, but the solutions approach weakly spatially homogeneous values although wildly oscillating in the strong sense of convergence. The regularization unveils the overall behavior of the solutions.


\section{Discussion}\label{sec.discussion}

\subsection{Relation to literature}
The Ehrenfest regularization (EhRe) was first presented in \cite{PKG-Landau}, where it was used to analyze the Ehrenfest reduction \cite{GK-Ehrenfest,GK-Ehrenfest2}. The regularization is in fact the first step of the Ehrenfest reduction, the next steps being projection to a reduced level of description, regularization of the projected reversible evolution and a closure. The regularization itself, however, is meaningful even on a single level of description, which is the subject of this paper.

The  combination of reversible Hamiltonian evolution and irreversible evolution (generated by a dissipation potential or dissipative matrix) was first developed in \cite{dv, MG84, grpd, MG85, Morrison}. Later if was called GENERIC and further developed in \cite{GO,OG}, see e.g. monographies \cite{hco,PKG}. Both the energetic and entropic regularizations have been shown to be compatible with GENERIC at least in the particular cases. The E-EhRe and S-EhRe can be thus seen as recipes for preparing the irreversible GENERIC evolution.

Double bracket dissipation, see \cite{Brockett1988, Bloch1990,Holm-Tronci2010} or \cite{Bloch-Krishnaprasad-Marsden1996}, is a neat geometric method for adding dissipation to Hamiltonian systems. A prototypical example is the equation for angular momentum with dissipation
\begin{equation}
	\dot{\mm} = \mm\times \oomega + \alpha \mm\times (\mm\times \oomega),
\end{equation}
which is compatible with dissipation in the Landau \& Lifschitz model \cite{LL1935}.
The double bracket dissipation is different from the Ehrenfest regularization, since the dissipative term contains only one derivative of energy while the EhRe contains two.

The double bracket dissipation moreover reduces energy of the system, which is not always the desired behavior in non-equilibrium thermodynamics of isolated systems, where energy is usually conserved and entropy (concave Casimir) produced. Moreover, when using the double bracket dissipation, little is usually told about evolution of entropy (concave Casimirs of the Poisson bracket). The Ehrenfest regularization can be seen advantageous by its possible forms (entropic, energetic or full), which can produce entropy and keep energy, reduce energy and keep entropy (Casimirs) or finally reduce energy and produce entropy. Moreover, the regularization is physically motivated by smoothing out solutions to the Hamiltonian equations.

Another interesting dissipative bracket was introduced in \cite{Holm-dissipative-bracket}, which can be seen as a generalization of the double bracket allowing for more general mobilities. Using Casimirs as generators, it can lead to selective Casimir decay \cite{Gay-Balmaz-Holm-decay} conserving energy and reducing Casimirs.

Geometric integrators (symplectic or Poisson) are being paid a lot of attention, see e.g. \cite{Leimkuehler,Hairer} and references therein. Although symplectic integrators were originally developed for ordinary differential equations (motion of particles, rigid body motion \cite{Simo1991}, etc.), the methods are proving useful also in partial differential equations, see e.g. \cite{Oers2017, Tang2017, Candy1991, Gagarina2016} or \cite{Gawlik}, where even Jacobi identity on the discrete level is recovered. Combination of reversible and irreversible dynamics with geometric integration was recently presented in \cite{HCO-integrator}, \cite{Materassi}, \cite{Morrison-plasma} (being essentially different from the approach in this manuscript). We would like to analyze symplecticity of the regularized equations with $\diff t = \tau$ in future.

\subsection{Non-convex energy}\label{sec.nonconvex}
Although it is not the main aim of this paper, let us answer the question what happens if we take the potential field non-convex, e.g.
\begin{equation}
	V(q) = V \cos(q/a),
\end{equation}
$V$ being an energy constant and $a$ a periodicity constant. Equations \eqref{eq.EhRe.qp} then become
\begin{subequations}
	\begin{eqnarray}
		\dot{q} &=& E_{p} + \frac{\tau}{2ma} V \sin(q/a)\\
		\dot{p} &=& -E_{q} +\frac{\tau}{2m a^2} V \cos(q/a) p.
	\end{eqnarray}
\end{subequations}
Considering for instance the particle near the local maximum at $q=0$, the irreversible term in the $\dot{q}$ equation pushes the particle out of the maximum. Moreover, the irreversible term in the $\dot{p}$ equation indicates growth of magnitude of the momentum. The local maximum of potential energy is thus unstable (as can also be checked by linearization of the equations easily). On the other hand, energy is convex near the local minima, which are thus stable in the sense above.


\section{Conclusion}
Hamiltonian evolution is purely reversible, but it can be, and often is, mathematically irregular. Such irregularities are then in a more overall viewpoint of the evolution smoothed out and the time evolution becomes irreversible and dissipative. 

Ehrenfest regularization (EhRe), Eq. \eqref{eq.x.evo.sr}, is a straightforward and thermodynamically consistent way of smoothing out the irregular solutions. The Hamiltonian vector field is followed for a short time. The new (regularized) vector generates  the time evolution made in small pieces of the original Hamiltonian trajectories.

The EhRe evolution reduces (kinetic, convex) energy while producing (kinetic) entropy, i.e. concave Casimirs of the Poisson bracket. By adding internal entropy, total energy (internal and kinetic) is conserved and internal entropy is produced. Moreover, the EhRe evolution can be discretized in time by forward Euler scheme with good properties. In particular, a second order explicit scheme for the reversible Hamiltonian equations can be obtained.

The EhRe evolution can be split into the energetic regularization (E-EhRe), Eq. \eqref{eq.EhRe.M}, and entropic regularization (S-EhRe), Eq. \eqref{eq.EhRe.N}. The former dissipates kinetic energy while preserving Casimirs of the Poisson bracket. The latter preserves kinetic energy while raising kinetic entropy (concave Casimirs). Both regularizations can be solved by Crank-Nicolson discretization in time, which keeps the desired properties.

EhRe, E-EhRe and S-EhRe are demonstrated on a particle in convex potential field (oscillator), rigid body motion, fluid mechanics and kinetic theory. In particular, the E-EhRe evolution indicates loss of stability of rotation around the minor axis of the rigid body, which is the sought physical behavior invisible in the original Hamiltonian equations for rigid body. In the case of fluid mechanics the best option seems to be to use the full EhRe evolution with internal entropy, which recovers energy and momentum conservation as well as growth of entropy. Alternatively, the full EhRe without internal entropy should lead to an explicit scheme for Euler equations for ideal fluids when taking $\diff t = \tau$. In the case of kinetic theory the S-EhRe evolution brings a new dissipative term to the evolution equation for the distribution function, which may lead to spatial homogenization (Landau damping).

In summary, having a Hamiltonian system, it is physically reasonable to alter the evolution equations by adding irreversible terms in order to manifest the overall behavior. If after adding the dissipation the system should conserve Casimirs of the underlying Poisson bracket (e.g. angular momentum), one should employ the E-EhRe method, which moreover reduces kinetic energy. If, on the other hand, kinetic energy is to be conserved, one should choose the S-EhRe, which moreover raises concave Casimirs of the bracket (kinetic entropy). Both E-EhRe and S-EhRe are discretized by Crank-Nicolson scheme. Or if neither kinetic energy nor Casimirs should be conserved, and energy should be dissipated while producing entropy, one should choose the EhRe evolution equations. In particular, if both energy and entropy should be conserved, one should choose the EhRe evolution with $\tau = \diff t$ and forward Euler scheme.

\section*{Acknowledgment}
We are grateful O{\u g}ul Esen, Josef M{\'a}lek and Tam{\' a}s F{\" u}l{\" o}p for many discussions encouraging the research.
This work was supported by Czech Science Foundation, project no.  17-15498Y, and by Charles University Research program No. UNCE/SCI/023.
This research has been supported partially by the Natural Sciences and Engineering Research Council of Canada, Grants 3100319 and 3100735.


\appendix

\section{Complete tangent lift}\label{sec.lift}
Consider a set of state variables $\xx$. Evolution equations of the state variables can in general be expressed as
\begin{equation}
	\dot{x}^i = v^i(\xx),
\end{equation}
where the dot stands for partial time-derivative and $v^i$ are components of a general vector field. The vector field on the space of state variables is then expressed as
\begin{equation}
	\vv = v^i \frac{\partial}{\partial x^i}.
\end{equation}

To the space (or manifold) of state variables one can always attach the tangent planes, and the tuple $(x^i,v^i)$ can be seen as an element of the tangent bundle. For instance vector field $\vv$ at point $\xx$ is an element of the tangent plane attached at $\xx$. The vector field can be lifted to a new vector field $\vv_c$ called complete tangent lift,
\begin{equation}
	\vv_c = v^i \frac{\partial}{\partial x^i} + v^j \frac{\partial v^i}{\partial x^j}\frac{\partial}{\partial v^i},
\end{equation}
that is a vector field on the tangent bundle, see e.g. \cite{Fecko}. This complete tangent lift expresses both how $\xx$ flows along the vector field $\vv$ and how the vector field $\vv$ itself changes along the flow.

The complete tangent lift is a vector field expressing the flow
\begin{subequations}
\begin{eqnarray}
	\dot{x}^i = v^i \\
	\dot{v}^i = v^j \frac{\partial v^i}{\partial x^j}.
\end{eqnarray}
\end{subequations}
After time $\tau$ the $v^i(t+\tau)$ variable will be approximately equal to
\begin{equation}
	v^i(t+\tau)\approx v^i(t) + \tau v^j \frac{\partial v^i}{\partial x^j}.
\end{equation}
Using the midpoint value $(v^i(t)+v^i(t+\tau))/2$ as the right hand side of the equation for $\xx$ leads to
\begin{equation}
	\frac{\diff x^i}{\diff \tau} = v^i(\xx(t)) + \frac{\tau}{2}v^j \frac{\partial v^i}{\partial x^j}\Big|_{\xx(t)},
\end{equation}
which is the regularized evolution of $\xx$ for a general vector field $\vv$. In particular, if $\vv$ is a Hamiltonian vector field, formula \eqref{eq.x.evo.sr} is recovered. This is the geometrical formulation of Ehrenfest regularization.

\section{Hamiltonian formulation of rigid body motion}\label{sec.Ham.SO3}

\subsection{Lie-Poisson bracket}\label{sec.LP}
Consider a rotating rigid body. The manifold of all possible states is the Lie group SO(3) of rotations in three-dimensional space. The Lie algebra associated to this Lie group (tangent space at the unit element of the group) consists of the infinitesimal rotation matrices, and its dimension is also three. Let us choose a basis of the Lie algebra,
\begin{equation}\label{eq.SO3.basis}
(\LL_i)_{jk} = -\eps_{ijk},
\end{equation}
i.e.
\begin{equation}
\LL_1 =
\begin{pmatrix}
0 & 0 & 0\\
0 & 0 & -1\\
0 & 1 & 0
\end{pmatrix},\,
\LL_2 =
\begin{pmatrix}
0 & 0 & 1\\
0 & 0 & 0\\
-1 & 0 & 0
\end{pmatrix},\,\mbox{and}\,
\LL_3 =
\begin{pmatrix}
0 & -1 & 0\\
1 & 0 & 0\\
0 & 0 & 0
\end{pmatrix}.
\end{equation}
Any element of the Lie algebra can be then expressed as $\XX = X^i \LL_i$. Commutators between elements of the basis are then
\begin{subequations}
\begin{eqnarray}
\left[\LL_1, \LL_2\right] &=& \LL_1\cdot\LL_2 - \LL_2\cdot\LL_1 = \LL_3\\
\left[\LL_2, \LL_3\right] &=& \LL_2\cdot\LL_3 - \LL_3\cdot\LL_2 = \LL_1\\
\left[\LL_3, \LL_1\right] &=& \LL_3\cdot\LL_1 - \LL_1\cdot\LL_3 = \LL_2,
\end{eqnarray}
\end{subequations}
which means that the Lie bracket of elements of the Lie algebra can be rewritten as
\begin{equation}
[\XX,\YY] = X^i [\LL_i, \LL_j] Y^j = \eps_{ijk}X^i Y^j \LL_k = \XX \times \YY,
\end{equation}
where $\XX$ and $\YY$ in the last expression are interpreted as vectors in $\R^3$ with components $X^i$ with respect to basis \eqref{eq.SO3.basis}.
The Lie algebra $\GGG$ of group SO(3) can be represented by vectors in $\R^3$ (components of $\XX$) equipped with the cross product.

When considering a general Lie group, the construction proceeds analogically as in the case of SO(3). The resulting Lie-Poisson bracket is
\begin{equation}\label{eq.PB.LP}
	\{F,G\} = -\langle \mmu [ F_\mmu, G_\mmu ] \rangle
\end{equation}
where $\langle\bullet,\bullet\rangle$ is a scalar product, $\mmu$ is an element of the Lie algebra dual and $[\bullet,\bullet]$ is the Lie bracket on the corresponding Lie algebra.

The dual space to $\R^3$ is again $\R^3$ and thus the Lie algebra dual $\GGG^*$ can also be represented by vectors in $\R^3$. Evolution of functionals on the dual, $F(\mm)$, $\mm\in \GGG^*$, is given by the Lie-Poisson bracket \eqref{eq.PB.LP}
\begin{equation}\label{eq.SO3.dF}
\dot{F} = - m_i \eps_{ijk} \frac{\partial F}{\partial m_j}\frac{\partial H}{\partial m_k}\stackrel{def}{=}\{F,H\}^{(SO(3))}
\end{equation}
where $H(\mm)$ is a yet unspecified Hamiltonian (or energy). Dynamics on the group of rotations in three-dimensional space SO(3) is thus generated by a Lie-Poisson bracket.

\end{document}